\documentclass[a4paper,fleqn]{cas-dc}
\usepackage{microtype}
\usepackage[numbers]{natbib}
\usepackage[dvipsnames]{xcolor}
\usepackage{array,multirow,makecell}
\usepackage{rotating}
\usepackage{graphics}
\usepackage{graphicx}
\usepackage{amssymb}
\usepackage{pifont}
\usepackage{caption}
\usepackage{subcaption}
\usepackage{inputenc}
\usepackage[colorinlistoftodos]{todonotes}
\usepackage{caption}
\usepackage{tabularx}
\usepackage{hyperref}
\usepackage{amsmath}
\usepackage{tabularx}
\usepackage{longtable}
\usepackage{float}
\floatstyle{plaintop}
\restylefloat{table}
\begin{document}
\let\WriteBookmarks\relax
\def\floatpagepagefraction{1}
\def\textpagefraction{.001}
\shorttitle{EEG based Major Depressive disorder and Bipolar disorder detection using Neural Networks: A review}
\shortauthors{Sana Yasin et~al.}

\title [mode = title]{EEG based Major Depressive disorder and Bipolar disorder detection using Neural Networks:A review}                      

\author[1,2]{Sana Yasin}
                        
\author[1]{ Syed Asad Hussain}     
   
\author[3,4]{ Sinem Aslan}    
   
\author[1]{ Imran Raza}   
   
\author[5]{ Muhammad Muzammel}

\author[5]{ Alice Othmani}
\cormark[1]

\address[1]{Department of Computer Science, COMSATS University Islamabad, Lahore Campus Lahore,Pakistan}
\address[2]{Department of Computer Science, University of Okara, Okara Pakistan}
\address[3]{Ca’ Foscari University of Venice, DAIS \& ECLT, Venice, Italy}
\address[4]{Ege University, International Computer Institute, Izmir, Turkey}
\address[5]{Université Paris-Est Créteil (UPEC), LISSI, Vitry sur Seine 94400, France}

\cortext[cor1]{Corresponding author: Associate professor. Dr. Alice OTHMANI \ead{alice.othmani@u-pec.fr}}

\begin{abstract}
Mental disorders represent critical public health challenges as they are leading contributors to the global burden of disease and intensely influence social and financial welfare of individuals. The present comprehensive review concentrate on the two mental disorders: Major depressive Disorder (MDD) and Bipolar Disorder (BD) with noteworthy publications during the last ten years. There is a big need nowadays for phenotypic characterization of psychiatric disorders with biomarkers. Electroencephalography (EEG) signals could offer a rich signature for MDD and BD and then they could improve understanding of pathophysiological mechanisms underling these mental disorders. In this review, we focus on the literature works adopting neural networks fed by EEG signals. Among those studies using EEG and neural networks, we have discussed a variety of EEG based protocols, biomarkers and public datasets for depression and bipolar disorder detection. 
We conclude with a discussion and valuable recommendations that will help to improve the reliability of developed models and for more accurate and more deterministic computational intelligence based systems in psychiatry. This review will prove to be a structured and valuable initial point for the researchers working on depression and bipolar disorders recognition by using EEG signals.
\end{abstract}

\begin{highlights}
\item A summary of clinical background of MDD and BD, Brain structure and EEG based brain activity,
\item EEG experimental protocols study for mental disorders diagnosis,
\item An extensive study of state-of-the-art shallow and deep neural networks models for EEG based clinical depression detection,
\item An extensive study of neural networks based approaches for Biploar Disorder diagnosis using EEG signals,
\item A discussion of state-of-the-art methods limitations and giving valuable recommendations for future research
\end{highlights}


\maketitle
\section{Introduction}
Mental disorders represent critical public health challenges because they are a leading contributors to the global burden of disease and profoundly impact social and economic welfare of the people. The World Health Organization predicted that by the year 2020, mental disorders shall be the first cause of disability worldwide
\footnote{\url{https://www.who.int/news-room/fact-sheets/detail/mental-disorders}}. According to WHO, it is predicted that over 264 million women and men of all ages bears some form of mental disorder, indicating that mental health problems effect up to 27\% of the general population at some point during their lives 
\footnote{\url{https://www.healthline.com/health/depression/effects-brain##1}}. The total cost of mental, neurological,  and substance use (MNS) disorders in the United States is over 210 billion, mostly related to indirect costs \cite{Ref183}. MNS disorders, therefore, account for 35\% of the overall burden of illness and are more costly than the combined burden of diabetes and cancers \cite{Ref184}. In contrast to other fields of medicine, psychiatry is still plagued by two problems: (1) a classification of mental disorders\footnote{International Classification of Diseases or the Diagnostic and Statistical Manual of Mental Disorders (DSM): \url{www.who.int/substance_abuse/terminology/diagnostic/en}} based on clinical symptoms of overlapping nosographic 
entities rather than on causal factors; (2) a pharmacologic arsenal that only targets clinical symptoms, mostly in an incomplete manner in a majority of patients. Moreover,
mapping diagnostic labels from a clinically defined nosology to varying biological indices has proven to be problematic \cite{Ref95}. Therefore, we need Mental Health Innovation and new ways to diagnose mental diseases by finding new biomarkers. Artificial Intelligence (AI) can play a key role in the psychiatry revolution. Multimodal Artificial Intelligence-based approaches and technologies need to be developed in order to advance our understanding and care of mental health, improve early precise diagnosis and prognosis, develop innovative treatments, and develop assistive technologies for longitudinal follow-up of the patient.

Recently, automatic recognition of mental states and mental disorders has attracted considerable attention from computer vision and artificial intelligence community. While, earlier works mostly adopted Functional Magnetic Resonance Imaging (fMRI) \cite{Ref128}, visual cues \cite{Ref127}, self-rating scale \cite{Ref9} and social network analysis \cite{Ref93}. 
In recent years non-invasive sensors based devices, such as Electroencephalogram (EEG), have been widely employed in the literature. One of the most remarkable research efforts has been made on developing efficient neural networks-based approaches for EEG signals analysis for automatic assessment of mental disorders such as Major Depressive Disorder (MDD) or Bipolar Disorder (BD).

Electroencephalogram (EEG) is a non-invasive, effective, and powerful tool for recording the electrical activity of the brain and for the diagnosis of various mental disorders such as MDD
\cite{Ref6}, BD \cite{vellante2020euthymic}, anxiety \cite{Ref148}, schizophrenia \cite{hebert2020electroretinogram}, and sleep disorders \cite{Ref149}. Due to these mental disorders or anomalies specifically depression and bipolar disorder, the body releases cortisol to the brain which affects the neurons production and communication and consequently slowing down the functionality of some parts of brain and changing the electrical activities patterns. The resulting voltage variations from ionic current flows within the neurons of the brain could help in the diagnosis of mental disorders like depression and bipolar disorder \cite{Ref30}. The development of robust approaches for brain signals analysis is a challenging task because of the complexity, the unstructured nature of signals and the big variability related to the person, to its age and its mental health. 
Moreover, brain signals are frequently affected by different types of noise due to eye blinking and muscular activities during EEG recording \cite{phadikar2020automatic}.

Major Depressive Disorder and Bipolar Disorder are mental disorders that affect physical health, sleep, appetite, attention level and they can lead to suicidal thoughts or actions. Both disorders have many similarities, but they also have some crucial differences. Patients with Bipolar disorder experience high mood swings that include excessive highs (mania or hypomania) and lows (depressive episodes). In mania episode, patients feel overactive, full of energy and usually irritable. While they feel opposite during depressive episodes with a strong feeling of hopeless and loss of pleasure in most of activities. Further, in depressive episodes they do not experience any elevated and extreme feelings that mania or hypomania patient face \cite{Ref139}.
 
The diagnosis of bipolar disorder is not always easy because many of the symptoms overlap with depression \cite{Ref124}.
Bipolar patients most probably consult their doctor for the first time when they have a depressive episode, instead of during a manic or hypo-manic episode. Due to this reason, clinicians frequently misdiagnose bipolar disorder as depression \cite{Ref124}.

In this paper, we present an exhaustive review of existing neural networks-based approaches, i.e., both shallow and deep architectures, for \emph{Major Depressive Disorder} (MDD) and \emph{Bipolar Disorder} (BD) diagnosis using EEG signals. With this review, we aim to provide a head start to the researchers with an up-to-date survey of advances of EEG based depression and bipolar disorders detection techniques for further contributions in this field. To the best of our knowledge, only a few surveys related to depression and EEG signals analysis, i.e., \cite{Ref5,Ref6,Ref7,Ref8,Ref9,Ref10}, have been published on depression diagnosis based on machine learning \cite{Ref5}, computer-aided diagnosis (CADx) \cite{Ref6}, speech analysis techniques\cite{Ref7}, nutrition examination \cite{Ref8}, self-rating scale \cite{Ref9} and by smartphone application \cite{Ref10}. Some of the surveys \cite{Ref96,Ref97,Ref98,Ref99,Ref100} cover the literature works on deep learning-based approaches using EEG signals. However, these techniques do not focus on depression diagnosis rather they consider other cognitive tasks \cite{Ref98,Ref99}, such as classification of brain signals \cite{Ref97}, motor imagery \cite{Ref100}, and Brain Computer Interfaces (BCI) \cite{Ref96}. To the best of our knowledge, this paper is the first comprehensive survey on neural network approaches that adopt either shallow or deep neural networks and using EEG signals for MDD and BD detection.

\subsection{Our Contributions} 
\label{sec:1.2}
In the past studies, deep learning and mental disorders fields have been studied separately. Just a few years ago, crossovers between these two areas have been merged and researchers have used deep learning for EEG-based mental disorders detection. Table ~\ref{tab:existing_surveys} shows the existing surveys related to deep learning, Electroencephalogram (EEG) and mental disorders.
To the best of our knowledge, this review is the first comprehensive study of the latest improvements and front lines of deep learning and artificial neural networks for MDD and BD recognition. In this survey, we have considered papers, most of which has been published in the last five years (since 2015). The contributions of this survey are as follows:
\begin{enumerate}
 \item A systematic review of artificial neural networks including shallow and deep learning-based approaches to detect Major depressive disorder (MDD) and Bipolar disorder using EEG, is presented to provide researchers an extensive understanding of this area of research.

\item Discussion on standard deep learning techniques and state-of-the-art models for MDD and BD detection, and providing some guidelines for choosing the suitable deep learning models.
\item A review of applications and challenges of deep learning and ANN-based MDD and BD detection. It also highlights some fascinating topics for future research.
\end{enumerate}

\begin{table*}
\caption{Summary of existing surveys and reviews related to DL, EEG,MDD and BD where DL stands for deep learning, EEG stands for Electroencephalogram, MDD stands for Major depressive disorder and BD stands for Bipolar disorder.} 
\begin{center}
\begin{tabularx}{12cm}{p{1.5cm}|p{1cm}|p{4cm}|X|X|X|X}
\hline Publication& Years & Topic &  \multicolumn{4}{c} {Scope} \\
    & & &DL&EEG &MDD &BD\\
    \hline
    \cite{Ref81} &2017 &  A survey on deep learning & \checkmark &x & x & x\\
    \cite{Ref82} & 2018 &A survey on deep learning & \checkmark &x & x & x\\
    \cite{Ref83} & 2019&A survey on deep learning & \checkmark & x & x & x \\
    \cite{Ref84} &2018& A survey on deep learning & \checkmark & x & x & x\\
    \cite{Ref85} & 2019 &A survey on deep learning & \checkmark & x & x & x\\
    \cite{Ref86} &2013& A survey on EEG   & x &\checkmark & x & x\\
    \cite{Ref87} & 2020 &A survey on EEG  & x &\checkmark & x & x\\
    \cite{Ref88}& 2020&A survey on EEG  & x &\checkmark & x& x \\
    \cite{Ref89}& 2019&A survey on EEG  & x &\checkmark & x& x\\
    \cite{Ref90} & 2020&A survey on EEG  & x &\checkmark & x& x \\
    \cite{Ref91}& 2014&A survey on depression  & x & x & \checkmark& x\\
    \cite{Ref92}& 2017&A survey on depression  & x & x & \checkmark& x \\
    \cite{Ref93}& 2020&A survey on depression & x & x& \checkmark& x \\
    \cite{Ref94} &2020& A survey on depression & x & x & \checkmark& x \\
    \cite{Ref95} & 2019&A survey on depression & x & x & \checkmark& x \\
    \cite{Ref96} & 2019&A survey on deep learning and EEG   & \checkmark & \checkmark & x& x \\
    \cite{Ref97}& 2018 &A survey on deep learning and EEG & \checkmark & \checkmark & x & x\\
    \cite{Ref98} & 2018&A survey on deep learning and EEG  & \checkmark & \checkmark & x& x \\
    \cite{Ref99} & 2019&A survey on deep learning and EEG  & \checkmark & \checkmark & x& x \\
    \cite{Ref100}& 2020&A survey on deep learning and EEG  & \checkmark & \checkmark & x& x \\
    \cite{Ref150}& 2020&A survey on bipolar disorder  & x & x & x& \checkmark \\
    \cite{Ref151}& 2020&A survey on depression and bipolar disorder  & x & x & \checkmark& \checkmark \\
     \cite{Ref152}& 2009&A survey on EEG based bipolar disorder  & x & \checkmark &x & \checkmark \\
    Our Review &2019&  Survey on EEG based MDD and BD detection using neural network & \checkmark & \checkmark & \checkmark & \checkmark \\
    \hline
\end{tabularx}
\end{center}
\label{tab:existing_surveys}
\end{table*}

\subsection{Survey Organization} 
\label{sec:1.4}

The current review is organized into nine sections. Section~\ref{sec:2} covers search strategy and eligibility criteria. Section~\ref{sec:3} describes the Electroencephalogram and brain structure. The section~\ref{sec:4} elaborates the clinical background of depression and summarizes the assessment of clinical depression by verbal and nonverbal signs. Section~\ref{sec:5} reviews the neural networks-based approaches, EEG experimental protocols and public datasets for depression recognition, while Section~\ref{sec:6} presents a clinical background of bipolar disorder by verbal and nonverbal signs. Section~\ref{sec:7} encloses the information about the neural networks-based approaches and EEG experimental protocols for bipolar disorder recognition. The discussion of the current review findings and suggestions for future studies are given in Section~ \ref{sec:8}. 
 At the end conclusion in section~\ref{sec:9} revealed the potential of shallow and deep neural network for depression and bipolar disorder recognition.

\section{Search Strategy and Eligibility Criteria}
\label{sec:2}

We have searched  IEEEXplore, PubMed, Embase, Springer, ScienceDirect and Web of Science, for articles published between January 2010, and May 2020 by using the following keywords: (“Depression” OR “shallow neural network” OR “Deep learning” OR “Electroencephalogram” OR “Cross-validation” OR “Bipolar depression” OR “Artificial neural network” OR “uni polar depression” OR “EEG base depression” OR “bipolar depression” OR “Major depressive disorder” OR “” OR “Recurrent neural network” OR “Deep neural network OR “BDI-II” OR “DSM-IV” OR “PHQ-9” OR “Persistent depressive disorder” OR “Deep Learning Models” OR “EEG bio-markers” OR “Deep Feature Extraction” OR “FFNN OR Feed forward neural network” OR “FBNN OR Feed backword neural network” OR “Discriminatives Deep learning models” OR “Representative Deep learning Models” OR “Generative Deep Learning Models” OR “Hybrid Deep learning Models” OR “Depression diagnosis's techniques” OR “Convolutional neural network OR CNN” OR “Mild Depression” OR “MLP” OR “Mental States” OR “ EEG Artifacts” OR “Bipolar Depression” OR “Manic Disorder” OR “Deep Belief Networks” OR “LSTM OR Long Short Term Memory. We also explored the articles that cite the ones that we found by the key-words mentioned above. 
There were no language restrictions.

We performed this systematic review by conforming to the PRISMA statement \cite{Ref131} that helps to improve the reporting of systematic reviews and meta-analyses. Eligibility criteria of this review includes the suitable depictions of different shallow and deep neural network techniques for the automatic assessment of depression by using EEG and representation of scientifically acquired data and generation of real-time results. Different technical reports and procedures \cite{Staples2007}, \cite{Strech2012} of systematic reviews are followed to complete this survey. Publication dates of the EEG based depression studies meeting the criteria of ten years is considered. Only those subjects are included in this survey that have more than 13 depression severity scores and no prior history of drug and medication. The main aim of this review is to sum up all the MDD and BD detection techniques that are performed by the shallow and neural network.
The Inclusion criteria for the current review encloses 1) adequate depiction of EEG and neural networks based automatic depression assessment classifiers and 2) demonstration of scientifically derived data, generating concrete and accurate results. Strictly clinical studies, as well as depression detection approaches relying on other techniques than EEG, are not included.
The keywords used to search electronic databases and related resources are listed in Table \ref{tab:keywords-2}. These keywords were used interchangeably, in combinations of two or more, with either “OR” or “AND” operands.

\begin{table*}
\caption{Keywords and web resources used in the current review.}
\begin{center}
\begin{tabularx}{13.8cm}{p{4.5cm}|p{2.8cm}|p{5.3cm}}
\hline
Web Resources & Theme & Keywords  \\
\hline
Elsevier & & Major Depressive Disorder\\
Springer & & MDD \\
Google Scholar & Depression & Clinical Depression\\
Mayo Clinic & & Unipolar disorder \\
PubMed & & Mild Depression\\
\cline{2-3}
Scopus & & BD\\
ACM Digital Library & & Mania\\
American Psychiatric Association & Bipolar Disorder& Manic depression\\
Embase & & Hypomania \\
NASA & & Manic-depressive illness\\
\cline{2-3}
Web of Science & & Shallow and deep neural network\\
IEEE Xplore Digital Library & &  FFNN or Feed forward neural network\\
World Health Organization & Deep Learning & FBNN  or  Feed  backword  neural  network\\
& & CNN or Convolutional neural network\\
& & RNN or Recurrent neural network \\ 
\cline{2-3}
&& EEG based depression\\
& Electroencephalogram & EEG-based unipolar depression\\
& & EEG based bipolar disorder\\
& & EEG biomarkers\\
\hline
\end{tabularx}
\end{center}
\label{tab:keywords-2}
\end{table*}

\section{Electroencephalogram (EEG) and Brain Structure: }
\label{sec:3}
Electroencephalogram (EEG) is an electro-biological measurement method that records the electrical activity of the brain signals that are highly random and encloses valuable information about the brain parts \cite{Ref134}. It records brain’s electrical activity over sometime by providing non-invasive and cost-effective solutions \cite{Ref143}. It is extensively used by physicians and researchers to study brain functions and to detect neurological syndromes. Due to the reliability factor of EEG, it is also utilized in EEG Biometrics for Person Verification \cite{Ref182}. Moreover, among different depression detection techniques (like audio \cite{Ref160}, facial \cite{Ref161}, text \cite{Ref162} and MRI \cite{Ref128}) EEG method is the most reliable due to its ease of use (i.e., it requires a simple placement of electrodes) and its high temporal resolution.

Depression and bipolar anomalies usually indicate dysfunction in the human brain. Abnormal shape of EEG signals appears as variations in the signals patterns for particular states of the patient and EEG reacts to the biotic activities of the brain, for the accurate detection of the brain abnormalities \cite{Ref143}. Both normal and depressed EEG signals are disordered and composite in nature with refined differences reflecting different brain activities of the depressive and normal groups that cannot be formulated easily through visual interpretations.

\subsection{Background of Electroencephalogram}
 The first human EEG recording was performed in 1924 by Hans Berger, who was a neuro psychiatrist at the University of Jena Germany. He gives the German name “elektrenkephalogramm” to EEG device that represents the graphical representation of the electric currents generated in the brain. Further, he presents that currents in the electrical pulses of the brain changes with respect to the functional status of the brain, such as, sleep, epilepsy and anesthesia. This idea of Hans Berger revolutionize the medical field and thus helped to open a new branch of medical science called neurophysiology. EEG signals can be divided into five different categories based on their bandwidth i.e., \textit{alpha} and \textit{beta}, \textit{theta}, \textit{delta} and \textit{gamma}, as illustrated in Fig \ref{fig:3}. Alpha and beta waves can be used to represent conscious states; while, theta and delta waves are mostly used to represent unconscious states \cite{sharma2012objective}. Gamma rhythm has been attributed to sensory perception \cite{nayak2019eeg}. The frequency ranges and major characteristics of brain waves are given in Table \ref{tab:Frequency ranges}.

  EEG signals can be recorded using invasive and non-invasive approaches. In an invasive approach, EEG electrodes are implanted in the brain by surgery or can be implanted at forehead sites. The manufacturing costs of these invasive electrodes are very high. While, the non-invasive electrodes are positioned on the scalp \cite{Ref143}. In a non-invasive approach, EEG signals can be recorded by using two types of electrodes, i.e., 1) wet electrodes and 2) dry electrodes. Wet electrodes are often made of silver chloride (AgCl) and  use a gel to create a conductive path between the electrodes and the skin by reducing the impedance value. The gel leakage can cause a short circuit between different electrodes. Furthermore, the extensive use of gel can cause allergy or any other infection. Therefore, non-invasive dry electrodes are also proposed to measure the EEG signals. These electrodes do not require gel or any skin penetration. Also, these electrodes work perfectly even on the hairy sites \cite{Ref31,Ref34,Ref38,Ref39,Ref40}.
 \begin{table*}[]
\caption{Electroencephalogram (EEG) bands and their characteristics.}
\begin{center}
 \begin{tabularx}{12cm}{p{2cm}|p{2cm}|p{3cm}|p{3cm}}
 \hline
 EEG Waves & Frequency Ranges(Hz) & Brain States& Mostly founded \\
 \hline
        Delta waves & 0.1-3Hz&Unconscious/Sleeping &Newborns and deep sleep phases\\
        Theta Waves & 4-8Hz  &Imagination&drowsiness and sleep\\
        Alpha waves & 8-13 Hz&Relaxed/Conscious &normal and relaxed subjects\\
        Beta waves& 13-30Hz&Conscious/Focused/ Problem solving&attentive or nervous subjects\\
        Gamma waves & 30-40Hz&Conscious perception/ Peak performance&Attentive subjects\\ 
        \hline
\end{tabularx}
\end{center}
     \label{tab:Frequency ranges}
 \end{table*}
  \subsection{Brain Structure}

\begin{figure}
    \centering
\begin{subfigure}{0.38\textwidth}
    \centering
    \includegraphics [width=1\linewidth]{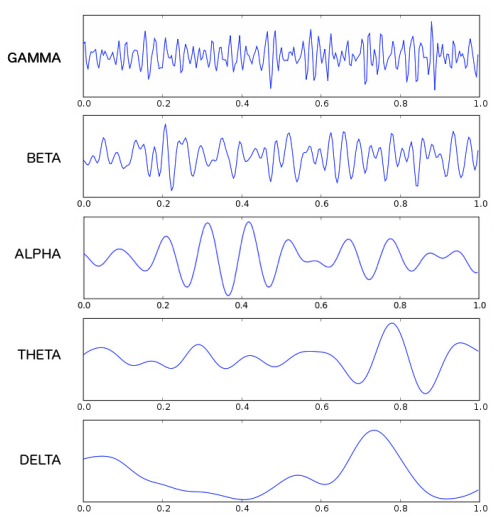}\caption{EEG waves (Adapted from\cite{Ref185}).}\label{fig:3}
    \end{subfigure}
    \begin{subfigure}{0.38\textwidth}
    \centering
    \includegraphics [width=1\linewidth]{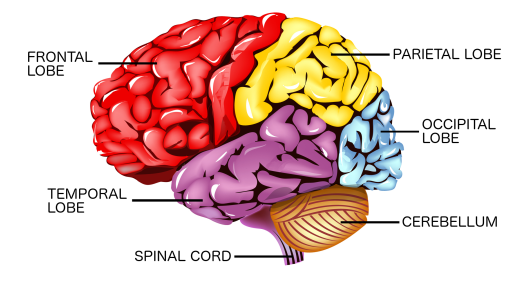}
    \caption{Representation of Brain Lobes (Adapted from \cite{Ref104}). }
    \label{fig:2}
	\end{subfigure}
	\begin{subfigure}{0.38\textwidth}
	\centering
    \includegraphics[width=1\linewidth]{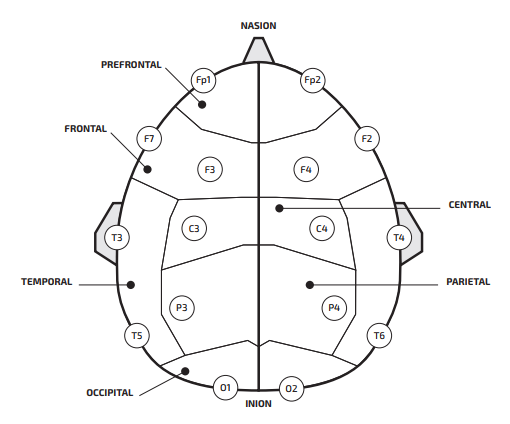}
    \caption{Birds eye view of electrode placement (Adapted from \cite{Ref47}).} 
    \label{fig:1}
\end{subfigure}
	
	\caption{a)Brain structure and EEG waves representation: b) Different brain lobes and their location; b) The placement of EEG electrodes at four brain lobes using 10-20 international system.} 
	\label{fig:aaaa}
\end{figure}

The Human brain is an incredible part of the body that controls all the body's functions and interprets the information from the outside world. It is composed of \textit{cerebrum}, \textit{cerebellum} and \textit{brainstem} that is enclosed within the skull. The cerebrum is a major part of the brain.  
It accomplishes complex functions like inferring touch, visualization, hearing, speaking, cognitive, sensations, learning, and adequate control of movement.
The cerebrum is made of left and right hemispheres, that controls the opposite side of the body and have distinctive fissures. Each hemisphere has four lobes: frontal, temporal, parietal, and occipital that are illustrated in Fig. \ref{fig:2}.

Each lobe represents different information of the human brain, i.e., frontal lobe controls the consciousness, the temporal lobe is responsible for the computation of complex stimuli and the senses of smell and sound, the parietal lobe represents the sensual information and the management of objects while the occipital lobe provides information about the sense of sight.
To extract the information about the depressive and non depressive subjects, EEG electrodes are placed at different lobes (frontal, temporal, parietal, and occipital) of the cortex, as can be seen in Fig. \ref{fig:1}.  
The placement of the electrodes at the scalp is important, because different lobes of the cerebral cortex are responsible for giving the information of electrical activities of the brain by mono polar and bipolar recordings \cite{Ref40}. Mono polar recording extracts the voltage variance among a reference electrode on the ear lobe and an electrode on the scalp. In contrast, Bipolar electrodes gathers the voltage variance among two scalp electrodes. 
\subsection{Effect of Depression and Bipolar Disorder on Brain Structure}
 
Depression and Bipolar Disorder affects three portions of the brain: \textit{hippocampus} (resides in the temporal lobe of the brain), \textit{prefrontal cortex} (located at the front of the frontal lobe) and \textit{amygdala} (the frontal portion of the temporal lobe) \cite{Ref165}. 
The hippocampus holds memories and controls the production of a hormone called cortisol. During depression body releases cortisol; that becomes problematic when its excessive amount is released and sent to the brain.
People with MDD, face long-term exposure of increased cortisol levels which can slow down the production of new neuron. It also causes the neurons in the hippocampus to shrink, hence leads to memory problems. 

The prefrontal cortex that resides in the frontal lobe is responsible for controlling emotions, making decisions, and creating memories. When the amount of cortisol exceeds in the brain, the prefrontal cortex gets shrink. The amygdala exists in the frontal portion of the temporal lobe and it enables emotional responses. 
In depression and bipolar disorder patients, the amygdala becomes large and more vigorous due to the continuous exposure of a high ratio of cortisol. An enlarged and manic amygdala, with irregular activity in other portions of the brain, can consequence in sleep disorders and other activity patterns. Usually, cortisol levels increase in the morning and reduce at night. However, in MDD patients cortisol ratio is always higher even at night. A Literature survey at EEG based depression and bipolar disorder detection represented in Table \ref{tab:electrode} of Section \ref{sec:3} shows that the frontal lobe is more effected by depression compared to the other brain lobes. 
\section{Clinical Background of Depression}
\label{sec:4}
Depression is a leading source of disability worldwide and significantly contributes to the global burden of disease. According to the World Health Organization (WHO)\footnote{\url{https://www.who.int/health-topics/depression}}, depression is a common psychiatric disorder characterized by a persistent undesirable effect like sadness, lack of attention, or pleasure in formerly satisfying or enjoyable activities. 
While the symptoms of the depression can be psychological (e.g., feeling hopeless, having continuous sadness, feelings of guilt and low self- esteem, loss of interest, having suicidal thoughts, and so on), the link between the depression and physical symptoms (e.g., headache, constipation, limb pain, stomach problems, pain at joints, back pain, tiredness, appetite and weight changes, sleep changes, and so on) has also been reported in the literature \cite{simon1999international,Ref144}.
The causes of depression comprise complex relations among psychosomatic, social, and biotic factors. Life moments such as childhood and teenage adversity, death, major events like losing a job, genetics, and substance abuse may increase the chance of depression. Pharmacological and psychosomatic treatments are available for moderate and severe depressive disorder. However, in low- and middle-income states, treatment and care facilities for depression are frequently absent or undersized \cite{Ref147}. WHO$^3$ reports that approximately 76–85\% of people in such countries have a lack of access to the depressive treatment they need.

According to Diagnostic and Statistical Manual of Mental Disorders (DSM-5) of the American Psychiatric Association (APA) \cite{Ref146}, depression exists in various forms like Major Depressive Disorder (MDD), Disruptive Mood Dysregulation Disorder (DMDD), Persistent Depressive Disorder (Dysthymia), Premenstrual Dysphoric Disorder (PDD), Substance/Medication-Induced Depressive Disorder (S/M-IDD), Depressive Disorder Due to Another Medical Condition (DDDAMC), and Other Specified Depressive Disorder (OSDD) or Unspecified Depressive Disorder (UDD). 

MDD, a.k.a. \textit{clinical depression}, is treated as the most typical form of the disease, and it is diagnosed by the existence of at least four of the following symptoms present for longer than two weeks \cite{Ref146,pampouchidou2017automatic}: changes in weight, changes in sleep, loss of energy almost every day, feelings of guilt and worthlessness, psychomotor agitation nearly every day, difficulty in concentrating, recurrent thoughts of death and suicide. In \cite{Ref164}, the etiology of MDD is associated with the genetic, biological, hormonal, immunological, neurological, environmental factors, acute life events and neuroendocrinological mechanisms.
While other types of depression have common symptoms with MDD, they mainly distinguish from MDD in a number of attributes. For example, DMDD is non episodic and occurs in children and teenagers. It is diagnosed by an obstinately ill-tempered, annoyed mood, and recurrent temper bursts. Dysthymia is an incessant and chronic form of depression. PDD causes severe irritability and nervousness in the week or two days before the period starts \cite{Ref137}. Substance/Medication-Induced Depressive Disorder (S/M-IDD) is a persistent form of MDD and occurs during or after substance intoxication. 
Depressive Disorder Due to Another Medical Condition (DDDAMC) is caused by long-term illnesses and ongoing pain in the body. 

In the following subsections we will be focusing on diagnosis and assessment of MDD, a.k.a clinical depression.
\subsection{Assessment of Clinical Depression}

Clinical Depression or MDD assessment includes an extensive  checkup of the patient,including examination of the mental state by discovering functional, relational, societal issues, and psychiatric history. Previous incidences of depression or mood rise,reaction to past treatment, and 
comorbid mental health disorders are examined. Regarding the level and effect of functional loss, or disability, clinical depression is assessed in mild, moderate or severe levels. Based on this assessment, the DSM-IV (Diagnostic and Statistical Manual of Mental Disorders) principles are used to make a diagnosis \cite{Ref130}, and the 
treatment is initiated. 
The benchmarks for diagnosing depression syndromes and their clinical consequences are set by the Diagnostic and Statistical Manual of Mental Disorders (DSM-5) \cite{american2013diagnostic}. 
In addition to a questionnaire-based approach where verbal signs are considered in depression-assessment, another followed approach is using biomarkers in which the biological process of the human body that is ongoing or has happened used as depression indication.

\subsubsection{Questionnaire-based assessment of Clinical Depression (Assessment by Verbal Signs)}
In this section, we highlight existing self-reported questioners-based depression detection techniques and their limitations. These questioners are designed for individuals and composed of questions to capture the 
signs of depression such as desperation, irritability, feeling guilty,  
and physical indications such as tiredness, weight loss, and lack of attention in sex, as well.  
Many instruments have been used by clinicians for depression assessment. Among these, we chose widely-accepted ones: PHQ-9, BDI, DSM-IV, CES-D, HAM-D. 

\textit{The Patient Health Questionnaire} (PHQ-9) is a multiple-choice self-report inventory and used to monitor the severity of depression \cite{kroenke2001phq}. It consists of nine items, each scored from 0 to 3, which results with a total score varying from 0 to 27. 
Based on the score range where the total score of the questioners fall into, subjects are categorized into different depression levels like Minimal, Mild, Moderate, Moderate severe, and Severe depression. Scoring system provided by PHQ-9 presented in Table \ref{tab:phq}.
In \cite{Ref43,Ref47,Ref49}, PHQ-9 is used for the participants' selection for EEG based experiments. 

\textit{Back Depression Inventory} (BDI) \cite{beck1996beck}, which is first published in 1961, is another questionnaire-based approach commonly used as a valuation tool by healthcare specialists and scientists to diagnose depression and anxiety. It has three versions BDI, BDI-1A and BDI-II. The first BDI was published in 1961 with 21 multiple choice questions against four possible responses of the patients.
BDI-1A, which is a revised version of BDI, was published in 1970. While BDI-1A provided ease of use, it still had some drawbacks, such as 
it only address 6 criteria of DSM-III out of 9. At the latest version, BDI-II, which was introduced in 1996,  
four items that are Body Image Change, Weight Loss, Somatic Preoccupation, and Work Difficulty, are changed by 
Agitation, Worthlessness, Concentration Difficulty, and Loss of Energy. Based on BDI-II,  
depression is classified into Minimal, Mild, Moderate, and Severe depression by the scoring system presented in Table \ref{tab:bdi}.   
In \cite{Ref33,Ref39,Ref40}, BDI is used as a psychometric test, i.e., 
every question has at least four possible responses, elongating in strength. A total high score shows the symptoms of severe depression. 

\textit{Diagnostic and Statistical Manual of Mental Disorders} (DSM-IV) \cite{american2013american}, which is published in 1994 by APA (American Psychiatric Association), has been used for the diagnosis of 
It offers a consistent classification system for the identification of mental health disorders for both adults and children. \cite{Ref38,Ref48,Ref54}, used DSM-IV to monitor   
perinatal depression patients for an EEG study.  
The updated version of DSM-IV was introduced in May 2013 with the title DSM-5 instead of DSM-V due to the limitation of the roman numerals \cite{Ref146}.  
The new version includes some essential
changes from the previous one, such as replacing the roman numbers to Arabic numbers, excluding Asperger's disorder and including disruptive mood dysregulation disorder.

The \textit{Center for Epidemiologic Studies Depression Scale} (CES-D) was initially designed for the general population. Since from last few years, it has been used for the screening of depression patients in primary care centers \cite{radloff1977self}. The CES-D contains 20 self-report questions, scored on a 4-point scale, which evaluate the type and level of depression experienced in the last few weeks. The CES-D can be applied to people of all ages. It has been verified across gender and cultural populations and achieve constant reliability and validity \cite{saracino2018confirmatory}.

The \textit{Hamilton Depression Rating Scale} (HAM-D) \cite{timmerby2017systematic} is a depression assessment tool that consists of  17 items that are used for scoring. 
The HAM-D emphasis on the wakefulness, desperateness, self-destructive thoughts, suicidal thoughts, and actions. It is chiefly used to diagnose the depression recovery in individuals before, during and after the treatment.

Questionnaire-based assessment approaches have several limitations. First, they are prone to professional's and patient's subjectivity, which hinders the objectivity of the process. Second, although depression presents different symptoms \cite{Ref102} and has high co-morbidity, especially with anxiety \cite{Ref103}, self questionnaire-based approaches are unable to evaluate differences across patient subgroups. Third, it is not capable of excluding participants already diagnosed as having or being treated for depression. Forth, it frequently performs a false diagnosis of bipolar disorder as a clinical depression \cite{Ref138}.
\begin{table}
 \caption{PHQ-9 scoring standard for depression detection (\cite{Ref156}).}
 \begin{tabular}{c|l} 
 \hline
  Depression Score & Depressive Severity \\
 \hline
   1-4 & Minimal Depression \\ 
   5-9 & Mild Depression  \\
  10-14 & Moderate Depression  \\
 15-19 & Moderately Severe Depression 	\\
 20-27 & Severe Depression\\
 \hline
\end{tabular}
     \label{tab:phq}
 \end{table}
 
\begin{table}
\caption{Back Depression Inventory BDI-II Scoring Standard for depression detection (\cite{Ref159}).}
\begin{tabular}{c|l}
 \hline 
 Depression Score & Depression Levels\\
 \hline 
 0–13 & minimal depression \\
14–19 & mild depression\\
20-28 & moderate depression\\
29-63 &Severe Depression\\
\hline
    \end{tabular}
    \label{tab:bdi}
\end{table}
\subsubsection{Biomarkers-based assessment of Clinical Depression (Assessment by Nonverbal Signs)}

A biomarker is defined as “a characteristic that is objectively measured and evaluated as an indication of normal biologic processes, pathogenic processes, or pharmacological responses to a therapeutic intervention” \cite{Ref101}. 
The Major Depressive Disorder (MDD) diagnosis and treatment can be enriched by decreasing the useless treatment trials 
by having accurate predictive bio-markers. Biomarkers-based assessment  provides accurate predictions of depression and bipolar disorder and is capable of adequately measuring the changes in disease conditions.
 
The scientific community uses different invasive and non-invasive tools and techniques as biomarkers to understand the mechanisms behind the MDD. 

The Anterior insula metabolism, the Hippocampal volume, and the Subcallosal cingulate cortex metabolism are \textit{neuroimaging-based biomarkers} 
\cite{Ref105}, and they are used as Treatment Selection Biomarker (TSB) for major depressive disorder. The limitation is that, these biomarkers may not be beneficial for long-term depression treatment selection, and may fail in severe depressive conditions.
Literature works report that genetic factors contribute to developing depression \cite{Ref178}. 
For example, it is shown in \cite{Ref177} 
that the gene 3p25-26 was found in more than 800 families with recurrent depression. 
\cite{Ref106} presents a review of EEG (Electroencephalograms) and ERP (Event-Related Potentials) based predictive biomarkers for MDD.  
This study highlights 1) Alpha power and asymmetry, 2) Theta band activations, 3) Antidepressant treatment response (ATR) index, 4) Theta QEEG cordance, 5) Referenced EEG (rEEG), 6) Rostral anterior cingulate cortex (rACC) activations and 7) machine learning as EEG based predictive biomarkers 8) The P300 \cite{Ref179} and LDAEP \cite{Ref180} as ERP biomarkers. 
Wake and sleep EEG is also used as a depression biomarker in \cite{Ref107} that provide an overview of sleep variations in depression.
It is reported in \cite{Ref181}
that 60–90\% of MDD patients with high severity suffers from sleep disorders.

In the past studies, while Alpha and Theta bands have been found to give information about the depression diagnosis and recovery 
\cite{Ref168,Ref111}, Gamma band was not well recognized in  
depression diagnosis \cite{Ref170}. 
On the other hand, in \cite{Ref109} Gamma waves are declared as depression diagnostic bio-marker by presenting some significant findings on gamma pulses.
In comparison to Alpha, Beta and Theta waves, Gamma waves have some distinct attributes: 1) Gamma pulses can accurately differentiate patients with major depressive disorder from healthy controls, under certain disorders, 2) Gamma waves can discriminate uni polar disorder from bipolar, 3) several pharmacological and no pharmacological treatments that counter depression also affect gamma. 
\cite{Ref110} adopts a variety of EEG features as depression diagnosis bio-markers extracted by linear and nonlinear methods and Phase Lagging Index (PLI) at the resting state of the patient. 

\cite{Ref111} used the frontal Theta asymmetry as a depression biomarker. More specifically, 
the EEG signals of 23 subjects with MDD and 23  
are recorded while they were listening to music. Results shows that frontal Theta power and frontal Theta asymmetry increased significantly in healthy subjects and decreased in depressed patients. 

\cite{Ref112} proposed to use multi-modal bio-markers (combination of executive dysfunctions, motor activity, neurophysiological patterns) for MDD diagnosis 
since depressive disorder affects not only mood but also psychomotor and cognitive functions. 20 MDD and 20 healthy subjects are selected. It is shown that the multi modal bio-markers are more consistent in the identification of MDD patients than the unimodal biomarkers are. 
\cite{Ref113} used brainwaves as a potential bio-marker for risk analysis of MDD. These brain waves 
are not only helpful in the depression diagnosis but it also provides the basic foundation for the accurate and reliable treatment of depression.

The EEG-based depressive bio-markers have several advantages over neuroimaging techniques \cite{Ref143}. While, the neuroimaging techniques are less effective and unable to provide information about the treatment, EEG bio-markers are easy to use, non-invasive, have high temporal resolution, cost-effective and provide optimal treatment selection. Despite lots of advantages of EEG-based bio markers, they have several drawbacks, such as  
poor measurement in below area of cortex and poor signal to noise ratio that require large number of participants for extracting useful information from EEG.
The major bio-markers that have been reported in the literature for depression diagnosis are shown in Table ~\ref{tab:biomarkers}.
\begin{table*}
\caption{ Psychological Biomarkers for the assessment of depression.}
\begin{center}
 \begin{tabularx}{15cm}{p{1cm}|p{2cm}|p{1.5cm}|p{2.5cm}|p{1.5cm}|p{4cm}}
 \hline
           Ref& EEG bio markers &ERP Bio markers& Neuroimaging Bio markers&  Genes and other Bio markers& Common Findings\\ [0.5ex] 
        \hline
       \cite{Ref168}&Alpha power and Asymmetry(FAA)&x& x & x & FAA is able to distinguish MDD and healthy control based on gender, age and depression severity.\\ 
            \hline
       \cite{Ref111}&Theta band activation's&x&x&x& Frontal theta asymmetry increased in normal subjects but reserved in depressed patients during music listening. \\ 
            \hline
         \cite{Ref169}&Antidepressant treatment response (ATR) index &x&x&x&ATR is a potential predictor and biomarker of NTR treatment of MDD pateints. \\ 
            \hline
        \cite{Ref170}&Theta QEEG cordance&x&x&x& The change in QEEG distingusish MDD and normal control.\\ 
            \hline
       \cite{Ref171} &Referenced EEG (rEEG) &x&x&x& It provides assistance in MDD medicine selection.\\ 
            \hline
        \cite{Ref173}&Rostral anterior cingulate cortex (rACC) &x&x&x& Low rAcc is a best responder to depression treatment.\\ 
            \hline
           \cite{Ref107,Ref108} &Wake and sleep EEG&x&x&x& The variation in wake and sleep helps in depression detection.  \\ 
            \hline 
            \cite{Ref109}& Gamma waves&x&x&x&gamma can discriminate uni polar disorder from bipolar. \\ 
            \hline 
        \cite{Ref172}&Resting state EEG&x &x&x& Extracted EEG features helps in depression diagnosis.\\ 
            \hline 
            \cite{Ref111}& Frontal theta Asymmetry&x&x&x& It provides reliable classification in comparison to gamma.\\ 
            \hline 
            \cite{Ref105}&x&x&Anterior insula metabolism&x&It identify treatment outcomes of depressed patients.\\
            \hline
          \cite{Ref175}&x&x&Psycho motor retardation&x& Affected by depression.\\ 
          \hline 
          \cite{Ref105}&x&x&Hippocampal volume&x&Depresses patient have 19\% small hippocampal. \\
            \hline 
            \cite{Ref174}&x&x& Cognitive functions&-&Cognitive actives gets damaged by depression.\\ 
            \hline 
            \cite{Ref10}&x&x& Subcallosal cingulate cortex metabolism&x&Abnormal SCG distinguish the MDD and control subjects.\\
            \hline
            \cite{Ref176}&x&x&x&Vocal and Facial bio markers&It is less reliable in comparison to EEG biomarker.\\
            \hline
            \cite{Ref177,Ref178}&x&x&x&3p25-26& This genes is found in more than 800 depressed  families but still immature.\\
            \hline
            \cite{Ref179}&x&P300&x&x& It is used as indicator to measure the severity of depression. \\
            \hline
            \cite{Ref180}&x&LDAEP&x&x& The higher LDAEP values found in depression patient.\\
            \hline
            \end{tabularx}
            \end{center}
     \label{tab:biomarkers}
 \end{table*}
 \section{Neural networks-based approaches for Depression recognition using EEG signals}
\label{sec:5}

The fields of Affective Computing (AC) and Neural Networks (NNs) are useful in solving complex and multidimensional problems such as modeling social affective behavior and mental disorders. These problems involving affective datasets need Deep Neural Networks (DNNs), neural networks with two or more hidden layers, for effective temporal modeling, and real-time performance analysis. The ability of DNN to identify latent structures in raw, unlabeled, unstructured, noisy, and incomplete EEG datasets makes it suitable for EEG based depression diagnosis. The effectiveness of DNN based solutions for depression diagnosis depends on the EEG experimental protocols, placement, and types of EEG electrodes, and the availability of EEG datasets\cite{othmani2019towards}. This also requires investigating automatic depression assessment methods. 

\subsection{EEG Experimental Protocols for Depression Recognition}
The EEG experimental protocol for depression recognition defines a standard set of rules such as number of participants, selection criteria of participants, EEG recording duration,and so on. Different EEG-based experimental protocols for depression recognition  
have been proposed in the literature. This section discusses the EEG based experimental protocol for depression recognition using shallow and deep neural networks.

\textbf{Participants:} The number of participants varies in different studies, from one to two hundred, with a median of 30 subjects. Most of these studies are based on a relatively small number of participants, therefore, it is hard to establish the exactness and significance of the results. Table \ref{tab:participants} presents a summary of existing studies highlighting the number of subjects, gender, age group, prior history, and depression. According to Table \ref{tab:participants},only ten out of fifty studies include thirty participants for the diagnosis of the EEG based depression. Thirteen studies include 50 to 60 subjects with an equal number of depressed and normal participants.The remaining studies consider different numbers of subjects.Regarding gender selection, most of the studies include both male and female participants; While, few studies include just female participants for EEG based major depressive disorder detection. Since men and women might perceive depression in different ways, so the participants from both genders must be included in studies.

\begin{table*}
\centering
\caption{Participants information in EEG experiments for depression detection where D for Depressed, H for healthy, F for female and M for male is used.}
 \begin{tabularx}{15cm}{p{1cm}|p{2cm}|p{1.5cm}|p{3cm}|p{1.2cm}|p{4cm}}
 \hline
 Ref & Subjects & Gender & Age Group\newline {(Mean, ± SD)} & Other Information & Depression Types\\
        \hline
            \cite{Ref31} & 63(33D+30H) & 27F,36M & 38-40 (38.22, 15.64) & - & Unipolar depression\\ 
        \hline
            \cite{Ref32} &30(15D+15H) & - & 20-50 & - & Bipolar depression \\
        \hline
            \cite{Ref33} &	51 & 15F, 36M & 18-24 (20.96, 1.95) & - & Mild depression \\
        \hline
            \cite{Ref34} & 30(15D+15H) & - &20-50 & - & Bipolar depression	\\
        \hline
                \cite{Ref35} & 60(30D+30H) & - & 20-50 & - & Bipolar depression\\
            \hline
                \cite{Ref36} & 10 & - & - & - & Depression\\
        \hline
                \cite{Ref37} & 60 & 30F, 30M & 10,32 (32.4, 10.5) & - & Minimal to severe depression\\
        \hline
                \cite{Ref38} & 28(14D+14H) & - & - & Right handed & Depression\\ 
        \hline
                \cite{Ref39} & 15 & - & - & - & Depression \\
                \hline
                \cite{Ref40} & 89 & - & - & - & Uni and Bipolar depression \\ 
        \hline
               \cite{Ref41} & 60(30D+30H) & 32F, 28M & 20-50 & - & Bipolar depression \\ 
               \hline
               \cite{Ref42} & 30 & 16F, 14M & 20-50 & - & Depression \\
        \hline
        \cite{Ref43} & 213(92D+121H) & - & - & -& Depression \\
        \hline
        \cite{Ref44} & 60(30D+30H) & 20-50 & - & -& Depression \\
      \hline
        \cite{Ref45} & 16 & - & - & - & Depression Level \\  
        \hline
         \cite{Ref46} & 51 & 15F, 36M & 18-24 & Right handed & Mild depression  \\ 
         \hline
         \cite{Ref47} & 116 & 50F, 66M & 19-25 & - & Mild depression  \\ 
         \hline
         \cite{Ref48} & 55 & - & - & - &  Major depressive disorder \\ 
         \hline
         \cite{Ref49}&34(17D+17H) &17F, 17M& 30-33 (33.35, 12.36)& - & Depression\\ 
         \hline
         \cite{Ref50} & 22 & - & - & - & Mild Depression  \\ 
         \hline
         \cite{Ref51} & 25(12D+13H) & 25F & 30-42 (24.23, 6.33) & Right handed & Pervasive depression\\
         \hline
         \cite{Ref53} & 178(86D+92H) & - & - & -& Depression\\
         \hline
         \cite{Ref54} & 24(12D+12H) &10F, 14M & 20-28 & - & Major Depressive Disorder \\
         \hline
         \cite{Ref55} & 30(15D+15H) & 16F, 14M& 20-50 & - & Depression\\
         \hline
       \cite{Ref59} & 30D & - & 20-50 & - & Depression \\
         \hline
         \cite{Ref60} & 64(34D+30H) & 26F, 38M & 15-38 & - & Major Depressive Disorder \\
         \hline
         \cite{Ref61} & 25(13D+12H) & 25F, 0M & (24.23, 6.33)& Right handed & Major depressive disorder \\
         \hline
         \cite{Ref62} & 23(12D+11H) & 12F, 11M &21-55 (26.4, 10.6) & - & Depression \\
         \hline
         \cite{Ref63} & 16 & - & - & -& Depression Level \\
         \hline
         \cite{Ref65} & 51(24D+27H) & 15F, 36M & 18-24 (20.96, 1.95) & - & Mild depression \\
         \hline
         \cite{Ref66} & 64(34D+30H) & 26F, 38M & 12-40 & - & Major depressive disorder \\
         \hline
         \cite{Ref67} & 12(12D) & 6F, 6M & 20-35& - & Depression\\
         \hline
         \cite{Ref69} & 30(10D+10H+10S) & - & 13-53 (31.5) & - & Depression \\
         \hline
        \cite{Ref70} & 22(12D+11H) & 10F, 12M & 20-24 & - & Depression \\
        \hline
        \cite{Ref71} & 20(10D+10H) & 10F, 10M & - & - & Depression \\
        \hline
        \cite{Ref72} & 116(63D+53H) & - & 19-25& - & Depression \\
        \hline
        \cite{Ref75} & 63(33D+30H) & 27F, 36M & 38-40 & - & Unipolar depression\\
        \hline
        \cite{Ref76} & 22(12D+10H)& -& 69-70 (69.81) & - & Depression\\
        \hline
 \end{tabularx}
     \label{tab:participants}
 \end{table*}
 \textbf{Selection Criteria of Participants:}
 The selection of the participants is the most important phase of EEG based depression recognition. The quantitative (clinical questionnaire-based) and qualitative (EEG-based) methods are used as standard tools for the selection of participants and data collection. In EEG based mental disorder studies, several clinical questionnaire-based pre-testing techniques are used in Table \ref{tab:selection} for participants’ recruitment, and then the patient’s selection for EEG recording.

Beck Depression Inventory (BDI) \cite{beck1996beck} is a common valuation tool used by healthcare specialists and researchers for pre-testing of depression and anxiety diagnosis. In this test, patients with MDD are selected based on different multiple-choice questions. The patients with BDI-II score above 13 are considered as a depressive subject. Overall high score of BDI-II shows the severity of depression. BDI is used in  \cite{Ref31,Ref33,Ref37,Ref39,Ref40,Ref41,Ref42,Ref44,Ref46,Ref61,Ref63,Ref64,Ref65,Ref75} for participants selection. Diagnostic and Statistical Manual of Mental Disorders (DSM-IV) developed by the American Psychiatric Association (APA) has been used by several studies for the measurement of different mental illnesses. In EEG based diagnosis, DSM-IV based questionnaire is used for pre-psychometric test to assess depression. Most of the articles used it as pre-EEG test \cite{Ref31,Ref38,Ref48,Ref60,Ref66}. Hospital Anxiety and Depression Scale (HADS) is used in \cite{Ref31} to measure the different levels of depression. Another participant selection method is known as PHQ-9 that is a multiple-choice self-report inventory. It consists of 9 questions and is used as a selection/diagnostic tool for mental health syndromes such as depression.

The subjects are categorized into different levels of depression such as minimal, mild, moderate, moderately severe, and severe depression based on the standard score of the questionnaire. The PHQ score standard is presented in Table \ref{tab:phq}. The most of the articles use it for pre-screening of the subjects \cite{Ref47,Ref49,Ref53,Ref72}. The participants are shortlisted for qualitative EEG based recording after the pre-screening test, and approval of the design of study from the ethics department. The participants perform specific tasks for a particular duration in a calm room under a resting state with a different number of non-invasive wet or dry electrodes at the different regions of the scalp. The recorded EEG signals contain a lot of noise that is removed for further processing of the signal and depression classification.
\begin{table*}
\caption{Participants selection and experimentation tasks in EEG studies for depression detection.}
\begin{center}
 \begin{tabularx}{15cm}{p{1cm}|p{3.5cm}|p{3cm}|p{2.5cm}|p{3cm}}
 \hline
  Ref & EEG Selection Criteria & Subject Inclusion criteria & Tasks& Depression criteria \\ [0.5ex] 
 \hline
  \cite{Ref31} & No psychotic disease &  DSM-IV & BDI,HADS &(BDI-II,HADS)score$>$14 \\ 
 \hline
 \cite{Ref32} & No medication & Psychiatrists interview & -& - \\ 
 \hline
 \cite{Ref33} & - & Investigation & Watch  pictures& BDI(II)score 14-28 \\ 
 \hline
 \cite{Ref34} & No medication & Psychiatrists interview & -& - \\ 
 \hline
 \cite{Ref35} & - & Psychiatrists interview & - & - \\ 
 \hline
 \cite{Ref36} & No mental disorder & interview & Working on ladder & - \\ 
 \hline
 \cite{Ref37} & No mental disorder & BDI-II & - & BDI(II)score 14-28   \\ 
 \hline
 \cite{Ref38} &  Free of neurological disease & MINI interview, & - & (DSM-IV) Score \\ 
 \hline
 \cite{Ref39} & No medication& -&Watch 5sec film of pictures & BDI(II), \newline{Hamilton score} \\ 
 \hline
 \cite{Ref40} & No mental disorder &-& Watch pictures & BDI(II)score 14-28 \\ 
 \hline
 \cite{Ref41} & No mental disorder & BDI-II & -&BDI(II)score 14-28 \\ 
 \hline
 \cite{Ref42} & No drug addiction & BDI(II) & Sit in rest position & BDI(II)score 14-28 \\ 
 \hline
 \cite{Ref43} & No medication & (PHQ-9), Mini-Mental State Exam(MMSE) & Sound stimulation & MMSE score \\ 
 \hline
 \cite{Ref44} & No epileptic & BDI(II) &-&score$>$14 \\
 \hline
 \cite{Ref45} & Free of drugs & - & - & - \\
 \hline
 \cite{Ref46} & No psychopathology & BDI-II & Watch picture &BDI-score$>$14 \\
 \hline
 \cite{Ref47} & No psychopathology & (PHQ)-9, DASS-21 & DASS-21& DASS-score$>$14 \\
 \hline
 \cite{Ref48} & No mood stabilizer & (DSM)IV,(HAM-D) & Resting position& HAM-D score$>$14 \\
 \hline
 \cite{Ref49} & Free of psychopathology &(PHQ)-9  & - &(PHQ)-score$>$5 \\
 \hline
 \cite{Ref50} & No psychopathic history & BDI-II &- & BDI-score$>$14 \\
 \hline
 \cite{Ref51} & Free of any medicine & BDI(II) & - & BDI(II)$>$13 \\
 \hline
 \cite{Ref42} & No neurological history & -  & - &- \\
 \hline
 \cite{Ref53} & Hospital inpatient & PSQI,PHQ-9,GAD-7,MINI  & Play sound track & PHQ-9$>$5 \\
 \hline
 \cite{Ref54} & No drug history & DSM-IV,(BDI)II  &- & (BDI)II-score$>$14 \\
 \hline
 \cite{Ref55} & No psychiatric drug & - & - & - \\
 \hline
 \cite{Ref59} & No mental disorder & Written consent&-&- \\
 \hline
 \cite{Ref60} & No mental disease & (DSM)-IV,HUSM  & - &(DSM)-score \\
 \hline
 \cite{Ref61} & Free of any medications & (BDI)II  &-& (BDI)II-score$>$14 \\
 \hline
 \cite{Ref62} & Free of any medications &Written consent&-&- \\
 \hline
 \cite{Ref63} & Free of drugs & (BDI)II  & - &(BDI)II-score$>$14 \\
 \hline
 \cite{Ref64} & No psychopathology. & (BDI)II  & Watch images & (BDI)II-score$>$14 \\
 \hline
 \cite{Ref65} &  No psychopathology & BDI-II & Watch pictures & (BDI)II-score$>$14 \\
 \hline
 \cite{Ref66} & No mental disease & DSM-IV & - & DSM-IV score \\
 \hline
 \cite{Ref67} & No medication & - & Play the VR game & - \\
 \hline
 \cite{Ref42} & No mental disease & Hospital patient & - & - \\
 \hline
 \cite{Ref69} & No psychiatric disease & Hospital patient & - &  - \\
 \hline
 \cite{Ref70} & No mental disease & Hospital patient &- &- \\
 \hline
 \cite{Ref71} & No mental disease & Hospital patient &- &- \\
 \hline
 \cite{Ref72} & No psychiatric disease & PHQ-9, DASS-21 & PHQ-9,DASS-21 &PHQ-9$>$5 \\
 \hline
 \cite{Ref75} & No psychotic &  DSM-IV & BDI,HADS & (BDI-II, HADS) score$>$14 \\
 \hline
 \cite{Ref76} & No psychotic &  MOCA & MMSE& MOCA score \\
 \hline
\end{tabularx}
\end{center}
     \label{tab:selection}
 \end{table*}

\textbf{Placement and Types of EEG Electrodes: }
The number of electrodes, their placement, and type play an important role in the EEG based depression and bipolar disorder diagnosis due to multiple reasons. First, the time required to set up the EEG device, second, ease of the patient who wears the EEG device, and third the number of features to process \cite{Ref28}.For these reasons, researchers have used different numbers of electrodes and standards to acquire EEG signals from the scalp. According to Table \ref{tab:electrode}, thirty one studies out of fifty use wet electrodes to record the EEG signals and the remaining use dry electrodes. The position of the electrodes at the scalp is also important because different lobes of the cerebral cortex are responsible for processing the electrical activities of the brain. As far as placement of electrodes at the scalp is concerned, two standards of 10-20 and 10-10 exist in the literature.

The 10-20 and 10-10 electrodes placement systems are based on an international standard that describes the location of electrodes at the scalp. In 10-20 standard, "10" and "20" represent the 10\% or 20\% inter-electrode distance and in 10-10 standard,"10" and "10" shows the 10\% inter-electrode distance. These standard electrode placement systems are based on the correlation between the location of an electrode and the underlying area of the cerebral cortex. The even numbers on the scalp represent the right hemisphere and odd numbers refer to the left hemisphere. The letters F, T, C, P, and O stand for Frontal, Temporal, Central, Parietal and Occipital respectively. They are used to identify the brain lobes and place the electrodes at the scalp. The point z refers to the midline of the brain. In the 10-20 electrode standard, the smallest number is closer to the midline and vice versa. A bird's eye view of electrode placement is shown in figure \ref{fig:1}.   According to Table \ref{tab:electrode}, twenty-five studies adopt 10-20 standard, only two use 10-10 standard, and remaining do not mention any information about electrode placement. According to Table \ref{tab:electrode}, number of electrodes varies in different studies. Researchers decide about the number of electrodes according to their requirements and diagnostic criteria. 
\begin{table*}
 \caption{EEG devices,number of channels and their placement used for EEG experiment for depression detection.}
 \begin{center}
    \begin{tabularx}{14cm}{p{1cm}|p{2.5cm}|p{3cm}|p{2cm}|p{1.5cm}|p{2cm}}
 \hline
  Ref & EEG Device & Electrode & Brain Lobes & Placement /Standard & Types of \newline{Electrodes}\\  
 \hline
  \cite{Ref31} & EEG cap & 19 &-&	10-20 &  Wet  \\ 
 \hline
 \cite{Ref32} & Pair electrode & 2 Channel pair &left/right half & - &  Wet  \\
 \hline
 \cite{Ref33} & (HCGSN) & 128 &-& 10-10 &  Dry  \\
 \hline
 \cite{Ref34} & (HCGSN) & 2 Channel pair &Left/right half& - &  Dry  \\
 \hline
 \cite{Ref35} & Bipolar Montage & 2 Channel pair &Frontal lobe& 10-20 &  Wet  \\
 \hline
 \cite{Ref36} & Salivary cortisol & 14 &-& - &  Wet  \\
 \hline
 \cite{Ref37} & - & 19 &-& - & Wet  \\
\hline
\cite{Ref38} & (HCGSN) & 16 &-& 10-20 &  Dry  \\ 
\hline
\cite{Ref39} & Procomp & 1(F4)&- & - &  Dry  \\
\hline
\cite{Ref40} & Procomp & 1(F4) &-& -  &  Dry \\
\hline
\cite{Ref41} & - & 24 & -&10-20 &  Wet \\
\hline
\cite{Ref42} & - & 2 Channel pair& Left/right half & - &  Wet \\
\hline
\cite{Ref43} & - & 3(Fp1,Fp2,Fpz) &Frontal lobe& 10-20 &  Wet\\
\hline
\cite{Ref44} & Bipolar Montage & 2 Chanel pair& Frontal lobe & 10-20 &  Wet\\[1ex]
\hline
\cite{Ref45} & Ag-Cl electrodes & 3 &Frontal lobe& 10-20 &  Wet \\
\hline
\cite{Ref46} & Geodesic,HCGSN & 16 &-& 10-20 & Dry \\
\hline
\cite{Ref47} & 32 Channel hardware &16& - & - &  Wet \\
\hline
\cite{Ref48} & (Compumedics/ Neuroscan) & 6&Frontal lobe & 10-20 &  Wet \\
\hline
\cite{Ref49} & HydroCel GSN (128)& 64 &-& 10-20 &  Dry \\
\hline
\cite{Ref50} & SynAmps & 61 & -&10-20 & Wet\\
\hline
\cite{Ref51} & Mobile EEG belt & 3(Fp1, Fp2 and Fpz)& Frontal lobe& 10-20 & Dry\\
\hline
\cite{Ref53} & - & 3(FP1,FP2,FPz) & Frontal lobe&10-20 &  Wet \\
\hline
\cite{Ref54} & - & 7(FP1-FP8) & Frontal lobe&10-20 &  Wet\\
\hline
\cite{Ref55} & - & 2 Channel pair & Left/right half&10-20 &  Wet \\
\hline
\cite{Ref59} & Bipolar Montage & 2 Channel pair & Left/right half &10-20 &  Wet\\
\hline
\cite{Ref60} & ProComm & 1(F4) &Frontal lobe & - &  Wet \\
\hline
\cite{Ref61} & EEG belt & 3 &Frontal lobe & 10-20 &  Wet \\
\hline
\cite{Ref62} & Brain amp & 6(Fp1, Fp2, F3, F4, P3 and P4)
 &Prefrontal,parietal cortex & 10-20 &  Wet \\
\hline
\cite{Ref63} & - & 3 & Frontal lobe &10-20 &  Wet \\
\hline
\cite{Ref64} & Geodesic sensor net (HCGSN) & 128& -& 10-20 &Dry \\
\hline
\cite{Ref65} & HCGSN & 128 & -&10-10 &  Dry\\
\hline
\cite{Ref66} & - & 19 &Temporal,parietal&10-20& Wet\\
\hline
\cite{Ref67} & -  & 3(Fp1, FpZ and Fp2) & Prefrontal lobe &10-20 &  Wet\\
\hline
\cite{Ref42} & -  & 2 Channel pair(FP1-T3)FP2-T4)&left/right half  & 10-20 &  Wet\\
\hline
\cite{Ref69} & HCGSN & 16(Fp1-FP8, C3-C4, P3-P4, T3-T6, O1-O2) &Frontal, ear lobe& 10-20 & Wet\\
\hline
\cite{Ref70} & -&5(F4, F3, Cz, M1, M2) &Frontal,earLobe& 10-20 &  Wet\\
\hline
\cite{Ref71} & - & 64 &-& 10-20 &  Wet\\
\hline
\cite{Ref72} & 32 channels wet EEG hardware & 16 &place at whole scalp& 10-20 &  Wet\\
\hline
 \cite{Ref75} & EEG cap & 19 &-&	10-20 &  Wet  \\ 
 \hline
 \cite{Ref76} & Nihon kohden JE-207A easy cap & 57 &-&	- &  Wet  \\ 
 \hline
\end{tabularx}
\end{center}
     \label{tab:electrode}
 \end{table*}
 \subsubsection{Public EEG Datasets for Depression Diagnosis}

 Due to the sensitive nature of depression data and for privacy and confidentiality reasons, very few public datasets are available for EEG based depression diagnosis, therefore, most research groups use their datasets. This section presents the few publicly available datasets for EEG based depression diagnoses as shown in Table \ref{tab:dataset}.
 
\textbf{Healthy Brain Network}\footnote{Healthy Brain Network Dataset link \url{http://fcon_1000.projects.nitrc.org/indG i/cmi_healthy_brain_network/sharing_neuro.html}}: HBN is a public biobank of data that is launched by the child mind institute. The major goal of HBN is to produce a dataset that captures a wide range of heterogeneity and impairment that occurs in evolving psychopathology. The HBN contains information about depressive disorder, behavioral, intellectual, eye tracking, and phenotype data, by using multi-modal EEG and brain imaging MRI. In the HBN, a clinical assessment of mental health and learning disorders is performed. The HBN protocols include the behavioral and physical measures, family structure, stress, trauma, cognition, and language tasks. About 10,000 subjects of New York with ages between 5 and 21years participated in data collection. The clinical team consisted of a mixture of psychologists and social workers. The safety and eligibility of the participants was confirmed by prescreening interview. The screening interview collected information about the subject’s psychiatric and medicinal history.

\textbf{EMBARC (Establishing Moderators/Mediators for a Biosignature of Antidepressant Response in Clinical Care)} 
is a public dataset made available by the National Institute of Mental Health (NIMH). It recognizes the neurological sign of reaction to antidepressant treatment by using the resting-state EEG and machine learning algorithm. A total of 16 posterior electrodes were comprised of P1, P2, P3, P4, P5, P6, P7, P8, PO3, PO4, PO7, PO8, POz, O1, O2, and Oz. \cite{Ref114}. 

\textbf{Depresjon} public dataset is used in \cite{Garcia:2018:NBP:3083187.3083216}, it contains motor activity recordings of 23 unipolar and bipolar depressed patients and 32 healthy controls\footnote{Depresjon Dataset link \url{https://datasets.simula.no/depresjon/\#data-collection}}.

\textbf{Trans diagnostic cohorts} \cite{Ref115} is a publicly available dataset that evaluates the brief transdiagnostic cognitive-behavioral group therapy (TCBGT) for the treatment of anxiety and depression patients. It contains 287 participants in primary care with depression and anxiety disorders. These patients spent approximately 5 weeks with TCBGT. ANOVA tests that have mixed design capabilities have been used for statistical analysis and to check the effects of treatment.

\textbf{Multimodal Resource for Studying Information Processing in the Developing Brain (MIPDB)} focuses on neuro-phenotyping of psychiatric and healthy populations of dimensional and multi-domains. They intellectualize mental disorder in terms of domain-general discrepancies, instead of considering a single factor\cite{Ref119}.

\textbf{Patient Repository for EEG data and computational tools (PREDICT)}
\footnote{Patient Repository for EEG data and computational tools (PREDICT) Dataset Link \url{Patient Repository for EEG data and computational tools (PREDICT}}
is a high volume publicly available dataset that contains EEG data. There exist several data repositories that contain imaging and patient-specific data. Patient Repository For EEG Data + Computational Tool (PRED + CT) is one of the few sites that offer EEG based patient-specific data. It provides a centralized platform by categorizing psychiatric and neural patients based on EEG data storage, tasks, and computational tools. In PRED+CT, EEG data implementation is performed in MATLAB toolbox based on patient/control, symptom scores, age, and sex \cite{Ref120}.
\begin{table*}
    \centering
     \caption{Public EEG datasets for depression diagnosis.}
    \begin{tabularx}{15cm}{p{1cm}|p{2cm}|p{1.5cm}|p{2cm}|p{2cm}|p{2.5cm}|p{1cm}}
    
\hline
     Ref& Dataset Name&Patient Group& Task during EEG recording&Number of Patients&EEG system&Electrodes  \\
     \hline
     \cite{Alexander2017}& Healthy Brain Network(HBN)& Depression&Resting state& 10,000& 128-channel EEG geodesic hydrocel system&128\\
     \hline
     \cite{Ref114}&EMBARC&Depression&Resting state&675&-&16\\
     \hline
     \cite{Garcia:2018:NBP:3083187.3083216}& Depresjon& Depression&Motor activity& 55(23D+32H)&-&-\\
     \hline
     \cite{Ref115}&Trans diagnostic cohorts&-&-&287&-&-\\
     \hline
     \cite{Ref116}&PREDICT (Depression Rest)&Depression /High BDI&Resting state&46&Neuroscan&64\\
     \hline
     \cite{Ref119}& MIPDB &Depressive&Resting state&126&-&109\\
     \hline
     \cite{Ref120}&PREDICT (Depression RL)& Depression RL/ High BDI&Reinforcement learning& 46&Neuroscan&64 \\
     \hline
     
\end{tabularx}
\label{tab:dataset}
\end{table*}

\subsection{Investigating Automatic Depression Assessment Methods}
In this section, automatic depression assessment methods are investigated by different prepossessing techniques, Neural networks (NNs), and deep learning-based approaches for depression detection.

\subsubsection{Pre-processing}
The EEG signal recording is a time-consuming procedure in which depressed patients perform some tasks. During recording, EEG signals are contaminated by undesired or polluted signals called artifacts. The artifacts that occur due to the patient body movement, heartbeat, eye blinks, muscle movement are called physiological artifacts. The artifacts that occur due to the electrodes placement, environmental noise, and device error are known as non-physiological artifacts \cite{Ref1}. These artifacts affect the quality of the actual EEG signals; hence it is important to sanitize useful data from contaminated EEG signals through pre processing of the EEG data.

In pre processing phase , different undesired artifacts are filtered by using different noise removal and artifact elimination algorithms to prepare data for the next stage. Table \ref{tab:preprocessing} shows the different artifact removal techniques. Mumtaz et al. \cite{Ref31}, claim that raw EEG signals have poor resolution due to the low Signal-to-Noise ratio (SNR). Therefore, to enhance the performance of EEG signals, the multiple source eye correction (MSEC) technique is used to remove the undesired signals. 

The authors of references  \cite{Ref32,Ref34,Ref44,Ref42,Ref55,Ref59,Ref67,Ref72} used a Notch filter with 50Hz to remove power line intervention and to sanitize EEG signals from artifacts. While, articles \cite{Ref33,Ref36,Ref37,Ref51,Ref55,Ref65,Ref66,Ref71,Ref72} used a low pass and high-pass-filter with 40Hz and 1Hz cutoff frequency to filter the noise. Adaptive noise canceller and fast ICA is used in \cite{Ref38,Ref76} to remove inaccurate information from the false EEG recordings.

As per our findings, the low pass, high pass, and Notch filter are frequently used to remove the artifacts in EEG based depression detection.
\begin{table*}
\caption{Artifacts and noise filtering approaches in EEG based depression detection. }
 \begin{center}
 \begin{tabularx}{9cm}{p{1cm}|p{8cm}}
 \hline
  Ref & Artifact filtering techniques \\ 
 \hline
  \cite{Ref31} & MSEC \\ 
 \hline
 \cite{Ref32} & Notch filter:50Hz  \\
 \hline
 \cite{Ref33} & High pass, net station waveform  \\
 \hline
 \cite{Ref34} & Notch filter:50HZ 	\\
 \hline
 \cite{Ref35} & -\\
 \hline
 \cite{Ref36} &Low-pass, high-pass, notch filter \\
 \hline
 \cite{Ref37} & Highpass, butterworth,low-pass \\
\hline
\cite{Ref38} & Adaptive noise canceller, Hzband-pass, fastICA \\ 
\hline
\cite{Ref39} & - \\ 
\hline
\cite{Ref40} & Band pass filter 40HZ \\ 
\hline
\cite{Ref41} & Total variation
Filtering (TVP) \\ 
\hline
\cite{Ref42} & Visual inspection, total variation filtering (TVF)\\
\hline
\cite{Ref43} & Kalman filter, adaptive predictor filter\\
\hline
\cite{Ref44} & Notch filter:50Hz \\[1ex] 
\hline
\cite{Ref45} & - \\
\hline
\cite{Ref46} & Net station waveform tool \\
\hline
\cite{Ref47} & High pass, Low pass,
and Notch filter \\
\hline
\cite{Ref48} & Band pass filter (0.15–30Hz) \\
\hline
\cite{Ref49} & FastICA, hanning filter \\
\hline
\cite{Ref50} & Off line ICA \\
\hline
\cite{Ref51} & Low pass filter at 50HZ \\
\hline
\cite{Ref42} & Notch filter at 50HZ \\
\hline
\cite{Ref53} & Band pass filter \\
\hline
\cite{Ref54} & Wavelet filter \\
\hline
\cite{Ref55} & Notch and low pass filter at 50HZ \\
\hline
\cite{Ref59} & Notch filter at 50HZ \\
\hline
\cite{Ref62} & Convolutional filter \\
\hline
\cite{Ref63} & ANFIS \\
\hline
\cite{Ref64} & Net station waveform tools \\
\hline
\cite{Ref65} & Low pass,high pass filter \\
\hline
\cite{Ref66} & ICA, notch, low pass,high pass filter \\
\hline
\cite{Ref67} & Notch filter \\
\hline
\cite{Ref42} & Visual inspection, 50-Hz notch filter \\
\hline
\cite{Ref69} & EEGLAB toolbox \\
\hline
\cite{Ref70} & Stationary wavelet transform (SWT) \\
\hline
\cite{Ref71} & Lowpass and highpass filter\\
\hline
\cite{Ref72} & Lowpass, highpass and notch filter\\
\hline
\cite{Ref75} & MSEC\\
\hline
\cite{Ref76} &ICA and butterworth filter \\
\hline
\end{tabularx}
\end{center}
\label{tab:preprocessing}
\end{table*}
\subsubsection{Neural networks-based approaches for depression recognition}

Neural Networks (NNs) \cite{jain1996artificial} are non-parametric, flexible, and parallel computing systems that consist of neurons layers and weighted links in which information is transferred from the input neurons to the output neurons in a forward or backward way. Neural Networks are broadly classified in to feed forward and feed backward network.

A feed forward neural network (FFNN) is developed by structured layers that are analogous to human brain neurons. Each layer consists of several connected units. The connection of units and layers are not uniform because each connection have different strength and weights. Example of FFNNs is the multilayer perceptron (MLP)\cite{Ref47}.

In recent years, ANN-based approaches are used in EEG studies for the classification and diagnosis of major depressive disorder. ANN can classify nonlinear relations and incorporates high-order dealings between predictive variables to produce accurate results. Most of the articles \cite{Ref40,Ref48} use it for the classification of unipolar and bipolar depression and have achieved good accuracy of 89.09\% by implementing the pre-treatment cordance of frontal QEEG.
 
Multi-layer FFNN, Back Propagation Neural Network (BPNN), and Enhanced probabilistic neural network (EPNN) have been used in \cite{Ref47,Ref51,Ref54} for discriminating MDD and non-MDD patients. In the test, EEG recording of depressed and age control subjects were collected under the 10 cross-validation technique. EEG electrodes were placed at the prefrontal (Fp1 and Fp2), frontal (F3, F4, F7, and F8), cerebral (C3 and C4), and temporal (T3, T4, T5 and T6) regions of the brain. The results indicate that the signals collected from the cerebral (C3 and C4) region gives slightly higher accuracy as compared to the other brain regions. Multi-layer FFNN performs better than BPNN and EPNN by achieving 95\% classification accuracy.

In \cite{Ref39,Ref41,Ref42}, a comparative study of FFNN, neuro-fuzzy networks, relative wavelet energy (RWE), and probabilistic neural networks (PNN) has been performed for differentiating the depressed and normal patients through EEG signals. Classification abilities of neural, neuro-fuzzy network, and relative wavelet energy (RWE) networks are authenticated by the EEG recordings. The FFNN leads the PNN, neuro-fuzzy networks, and relative wavelet energy (RWE) achieving 100\% classification accuracy with time and wavelet energy as the input features.

EEG entropies based depression detection has been performed in \cite{Ref55,Ref56} by comparing the Probabilistic Neural Network (PNN) performance with Radial Basis Function (RBF) and different machine learning classifiers. Results shows that PNN performs better with 99.5\% classification accuracy.

\subsubsection{Deep learning based approaches for depression detection}

Deep learning is the most popular area of research and currently, it is extensively used for the classification of EEG signals in comparison to other approaches\cite{bachmann2018methods}.
Deep learning gives accurate predictions,in EEG based depression classification, than traditional machine learning due to the huge volume of data for training. In comparison to the traditional Artificial Neural Network (ANN), that are trained by a supervised algorithm with limited to few layers, deep learning uses multiple layers to extract complex features from the raw EEG data \cite{muzammel2020audvowelconsnet}. 

The convolutional neural network (CNN), long Short-Term Memory (LSTM), recurrent neural network (RNN), denoising autoencoder (DAE) and deep belief networks (DBNs) are the popular deep learning methods that are widely used in most of the EEG studies to diagnose major depressive disorder\cite{mosavi2019list}. The performance comparison of all these methods in existing EEG based depression detection studies are shows in Table \ref{tab:classifiers}. 

\begin{figure*}
    \centering
    \includegraphics{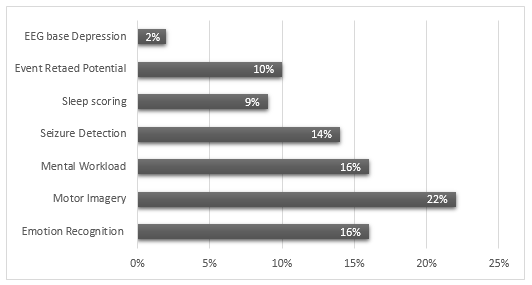}
    \caption{Ratio of Deep Learning studies adopting EEG signals for a variety of applications referred to \cite{Ref117}.}
    \label{fig:deep learning classification}
\end{figure*}

A literature survey shows that approximately there are six different areas of EEG: sleep, seizure prediction, mental workload, motor imagery, emotion recognition and depression that use deep learning for the detection and classification purposes \cite{Ref117}. According to the latest survey on deep learning \cite{Ref117}, it is observed that there are 16\% articles for EEG based emotion recognition system, 22\% for motor imagery, 16\% for mental workload, 14\% for seizure detection, 9\% for sleep stage scoring, 10\% of event-related potential detection, 2\%  for EEG based depression and 8\% for Alzheimer’s, gender and abnormal signal classification(Please see Fig. \ref{fig:deep learning classification}).

Herein, we see that the motor imagery is the most frequently explored area but the use of deep learning in EEG based depression recognition is not much reported so far. 
In \cite{Ref31} authors claim that there are several articles on the automatic diagnosis of depression with traditional classifiers, however, there are only two or three articles on deep learning methods for EEG based depression detection.

In the existing studies, questionnaires \cite{Ref92}, visual cues \cite{Ref127}, regression and machine learning-based techniques \cite{Ref5} are used for the depression recognition and treatment. Besides all these efforts, the depression detection field still needs improvements. To enhance the feature extraction and classification accuracy for depression recognition and assessment tasks, deep learning is used for the last two years.

The Convolutional Neural Network (CNN) is the most famous model of deep learning and is widely used in most of EEG studies to diagnose major depressive disorder. One Dimensional Convolutional Neural Network (1DCNN) is used with Long Short Term Memory (LSTM) for the classification of unipolar depression \cite{Ref31}. This hybrid architecture achieves 95.97\% accuracy by automatically learning EEG patterns.
  The cerebral cortex is divided into the left and right hemispheres of the brain. In \cite{Ref32}, the authors used Deep Convolutional Neural Networks to classify depressive and non-depressive subjects against the left and right hemispheres. Results shows that the right hemisphere is more accurate in depression classification as compared to the left hemisphere with 93.5\% accuracy. Mild depression recognition of college students has been performed \cite{Ref33} by the convnets framework.

  The convnets framework achieves 85.62\% accuracy with 24-fold cross-validation. Automatic depression detection using deep representation and sequence learning is performed in \cite{Ref34}. These two hybrid architectures of deep learning provide sequence learning and extract the temporal properties of the EEG signals. It achieves 97.66\% classification accuracy by placing the electrodes at the left and the right hemisphere of the brain.

The Neucube architecture of deep learning with 10 cross-validations is applied in \cite{Ref50} with the resting state of EEG. The neucube architecture is compared with Multi-layer Perceptron (MLP) and other traditional machine learning methods. The results shows that neucube performs better than MLP and all other traditional machine learning algorithms such as SVM, decision tree and logistic regression. Deep learning-based depression detection with frontal imagery EEG channels is performed in \cite{Ref75}. The Oxford Net (VGG16) model with $3\times 3$ max pooling and softmax activation function is used. The results shows that OxfordNet (VGG16) has 87.5\% classification accuracy. Functional connectivity and EEG based mild depression detection has been performed in \cite{Ref78} by using a deep learning approach. Initially, abnormal functional connectivity is measured by using the graph theory, and then CNN is applied for the classification of the depression with 80.74\% accuracy.

Deep neural network approaches have also been used for feature extraction for transformation of the huge volume of input data into imperative features sets. In Major Depressive Disorder (MDD), the feature extraction phase selects the most imperative features or information from the EEG signals for depression classification. The feature extraction part has a great influence on the accuracy of the results.

In \cite{Ref31} temporal feature is extracted efficiently by using the One Dimensional Convolutional Network (1DCNN). The first layer of the CNN layer maps the 64 features, the next layer reduced it from 64 to 48, and the third layer includes 24 features only by using the pooling layer and gets better experimental results. In \cite{Ref32} authors do not use any manual set of features to be fed into a depression classifier. The given CNN model has the capability to self-learn and select the unique features during the training process without a separate feature selection step.

In \cite{Ref33} pre-trained model based on transfer learning is used to extract the power spectrum of EEG signals observing the differences between depressed and non-depressed patients in alpha, beta, and theta frequency bands. The pre-trained model removes the last output layer of CNN and then uses the entire network as a fixed feature extractor. In \cite{Ref34} both local and long term feature selection/extraction and depression classification operations are automatically performed by using an end-to-end single framework and EEG signals as input. Convolutional and pooling layers of deep learning models are used in \cite{Ref36} for feature extraction. The benefit of these layers is that they enhance the classification accuracy by extracting the features from the nearest neighboring pixels. Brain has a nonlinear and complex system so nonlinear features including fuzzy entropy FuzzyEn and fractal dimensions (KFD and FFD) are used in \cite{Ref37}.

Ensemble learning and deep learning approaches are applied in \cite{Ref38} for the processing of features from brain signals, power spectral density and activity are extracted as original features. Neucube model as a feature extractor is applied in \cite{Ref50}. It achieves not only high classification accuracy, but also reveals patterns of brain activities relevant to the two classes of depressive and non-depressive subjects.

The performances of a variety of EEG based deep learning and neural network approaches for depression detection are presented in Table \ref{tab:classifiers}.
\begin{table*}[]
\caption{Performance of different EEG based deep learning and neural network approaches for depression detection presentation in terms of accuracy. 
}
 \centering
 {\begin{tabularx}{12cm}{p{1cm}|p{7cm}|p{4cm}}
 \hline
  Ref& Model used & Results(Accuracy) \\ [0.5ex] 
 \hline
  \cite{Ref31}& 1DCNN+LSTM based on repeated blocks of (C1-P1-DP1) layers &	98.32\% \\ 
 \hline
 \cite{Ref32}& 13 Layer of
Convolutional neural network& 95.49\% \\
 \hline
 \cite{Ref33}& Computer-aided design using convNet& 85.62\% \\
 \hline
 \cite{Ref34}&Hybrid model (CNN+LSTM) & 99.12\% \\
 \hline
 \cite{Ref35}& 3 layer Convolutional neural network & 99.31\%	  \\
 \hline
 \cite{Ref36}& CNN and Fully Connected Neural Network& 86.62\%\\
 \hline
 \cite{Ref37}& Fuzzy neural network (FFNN) & 87.5\%\\
\hline
\cite{Ref38}& 7 layer CNN and SVM & 84.75\% \\
\hline
\cite{Ref39}& 3 layer NN and Neuro-Fuzzy network &13.97\% and76.88\% \\
\hline
\cite{Ref40}& FFNN-Particle Swarm Optimization& 89.89\%  \\
\hline
\cite{Ref41}&Feedforward, probabilistic neural network & 58.75\% and 98.75\% \\
\hline
\cite{Ref42}& Relative wavelet energy (RWE) and artificial feedForward neural network &  98.11\% \\
\hline
\cite{Ref43}& SVM, KNN, CT, ANN &  72.56\% \\
\hline 
\cite{Ref44}& 5 layer CNN & 99.3\%\\
\hline
\cite{Ref45}& Neural network pattern recognition tool (nprtool)
and ANFIS tool box  &  91.7\% \\
\hline 
\cite{Ref46}&Self Normalizing Neural Networks (SNN) & 83.42\% \\
\hline
\cite{Ref47}&3 distinct layer feed-forward neural network & 95\% \\
\hline
\cite{Ref48}& Artificial Neural Network with 5 neurons & 89.09\% \\
\hline
\cite{Ref49}& 4-layer ConvNet & 77.20\% \\
\hline
\cite{Ref50}& SNN NeuCube architecture. & 90\% \\
\hline
\cite{Ref51}& 
Back-Propagation Neural Networks (BPNN) & 94.2\% \\
\hline
\cite{Ref42}& Feed forward neural network & 98.11\% \\
\hline
\cite{Ref53}& Artificial Neural Network, Deep Belief Networks & 78.24\% \\
\hline
\cite{Ref54}& Enhanced probabilistic neural network  & 91.3 \%\\[1ex]
\hline
\cite{Ref55}& Probabilistic neural network & 99.5\% \\[1ex]
\hline
\cite{Ref56}& Artificial neural network with 20 neurons & -\\
\hline
\cite{Ref58}& Multilayer perceptron &80\%\\
\hline
\cite{Ref59}& Recurrent neural network,Long short term memory & 80\%\\
\hline
\cite{Ref60}& 
Multiple-layer perceptron neural network & 93.33\% \\
\hline
\cite{Ref61}& Back-Propagation Neural Networks& 94.2\% \\
\hline
\cite{Ref62}& Hybrid EEGNet& 79.08\%\\
\hline
\cite{Ref63}& ANFIS, NPR (neural network tool) & 91.7\% \\[1ex]
\hline
\cite{Ref64}& ANN (BNMLP) & 83.42\% \\[1ex]
\hline
\cite{Ref65}& Artificial neural network& 85.62\%\\
\hline
\cite{Ref66}& Multilayer perceptron neural network & 93.33\%\\
\hline
\cite{Ref67}& FBNN & 70\%\\
\hline
\cite{Ref42}& Artificial neural network with 10 neurons& 98.11\%\\
\hline
\cite{Ref69}& Artificial neural network with one hidden layer & -\\
\hline
\cite{Ref70}&Back-Propagation Neural Networks & -\\
\hline
\cite{Ref71}& Artificial neural network& -\\
\hline
\cite{Ref72}&  Artificial neural network& 95\%\\
\hline
\cite{Ref73}& Probabilistic neural network& 98\%\\
\hline
\cite{Ref74}&Deep learning (SPN) & -\\
\hline
\cite{Ref75}&Multilayer Convulational neural network & 87.5\%\\
\hline
\cite{Ref76}& Multilayer perceptron (MLP) & 95.45\%\\
\hline
\cite{Ref77}& Stochastic gradient descent algorithm with deep neural network & 95.45\%\\
\hline
\cite{Ref78}&  Two stacked convolutional neural networks & 80.74\%\\
\hline
\end{tabularx}}
\label{tab:classifiers}
\end{table*}
\section{Clinical background of Bipolar Disorder}
\label{sec:6}
Bipolar Disorder (BD) or manic-depressive illness (MDI) is a dangerous neural disorder that is predicted by mood uncertainty and may start from early ages (i.e., in infants and teenagers). Individuals with BD face swings between depressive and manic episodes recurrently \cite{Ref139}. According to the American Psychiatric Association \cite{Ref124}, there are four major categories of bipolar disorder: \textbf{bipolar I disorder}, that has one atleast one full episode of mania or diverse episodes of mania and depression.\textbf{ Bipolar II disorder} has no manic episode, minimum one hypomanic episode and many depressive episodes. Patients with \textbf{cyclothymic disorder} have 
many hypomanic and depressive episodes. \textbf{Bipolar disorder} includes both depressive and hypomanic episodes alternatively.\\

  Bipolar disorder and depression have various resemblances, yet they have some crucial dissimilarities \cite{Ref142}. Patients with bipolar disorder have experience of high mood swings, different episodes of depression and periods of excessive highs (also known as mania). In contrast, depression is just one state of bipolar disorder that is more than from the low feeling. It's a deep grief or hollowness that an individual cannot manage. Sometimes, bipolar disorder patients feel disheartened, valueless, tired and may fail to concentrate on activities that they had enjoyed before the illness. Clinical depression (also known as major depressive disorder or MDD) frequently comes with sleep difficulties, and alterations in hunger. It can lead to suicidal activities or actions. People with major depressive disorder do not practice any extreme, elevated feelings that bipolar patients patients face with mania or hypomania.

In bipolar disorder, the symptoms of intense periods with low and high moods do not follow a pattern. Some bipolar patients experience the same mood state for a long period before shifting to the opposite mood and vise versa \cite{samalin2014patients}. These depressive periods occur for a week, month, or sometimes even for a year. Bipolar severity varies with respect to person and occurrence time. Bipolar disorder symptoms are broadly classified into two types: 1) External and physical, and ii) Internal and physiological symptoms.

Bipolar disorder has a strong physical effect on patient's body and creates many physical problems of flu, palpitations, diarrhea, abdominal pain, nausea and vomiting, high pulse and heart rates, higher blood pressure, weight and appetite changes, fast speaking and poor attention, strangely high sex drive, enhanced energy and less need for sleep \cite{Ref166,Ref167}.
 The primary physiological symptoms of bipolar disorder are penetrating and unpredictable. It includes severe mood swings, extreme pleasure, hopefulness and excitement, rapid changes from being glad to being ill-tempered, irritated, aggressive and becoming more thoughtless. In the following subsections we will focus on diagnosis and assessments of clinical bipolar disorder.

 Irregular physical features of brain or an inequality in certain brain chemicals are the main causes of bipolar disorder. The assessment of bipolar disorder is not always easy like other mental disorders \cite{Ref140}. A bipolar patient most probably goes to the consultant for the first time when he/she has a depressive episode instead of during a manic or hypomanic episode. Due to this reason, in the beginning clinicians frequently misdiagnose bipolar disorder as depression. By considering all limitations of bipolar disorder, scientists and researchers have introduced some automatic and reliable sources for the diagnosis and treatment of bipolar disorder. Nothing is more significant than diagnosing a patient with bipolar disorder or manic-depressive illness (MDI), as only accurate diagnosis can lead to a proper effective treatment. Electroencephalogram (EEG) is a top rated neuroimaging technique that is becoming a central focus of researchers for the past few years and is widely used to diagnose mental disorders.
 
\subsection{Questionnaire-based assessment of Bipolar disorder (Assessment by Verbal Signs)}
Like major depressive disorders, questionnaire based assessment tools are also used for bipolar disorder participants to investigate their physiological responses. The Young Mania Rating Scale (YMRS), Hamilton Depression Rating Scale (HDRS) and Structured Clinical Interview for Diagnostic (SCID), are the few bipolar assessment tools that are widely used for participants selection \cite{Ref40}. The YMRS \cite{young2000young} is one of the most widely used rating scale to evaluate the mania symptoms. It consists of 11 items to evaluate the clinical condition of patients in the last 48 hours. The purpose of each item is to measure the abnormality of the patient. After this, HDRS is used to measure the severity of the disease. 
Based on the clinical feature and symptoms, patients are categorized into different diseases e.g., YMRS=12 show bipolar disorder, YMRS=3 shows depression and YMRS=2 shows ethymia.

The HDRS \cite{williams1988structured} was originally published in 1960 and was used to evaluate the recovery process (and also to measure the severity) of major depressive and bipolar disorders. It contains ten multiple choice based questions that provide an indication to bipolar depression. It usually works for adults and measure the severity level of depression and mania according to the HDRS total score. 
The SCID \cite{first2014structured} is a structured diagnostic tool developed in 1990 and it works with different versions of DSM to determine their axis. It is organized into different modules and each module is used to detect unique type of disorder i.e., SCID-I diagnose mental disorders, SCID-II  determine personality disorders and SCID-5 diagnose the anxiety, eating, gambling and sleep disorders.

\subsection{Bio markers based assessment of Bipolar disorder (Assessment by Nonverbal Signs)}
The pathophysiology of bipolar disorder is complex, multi factorial, and not fully understandable. To overcome this complexity, the biomarkers based assessment not only facilitates the diagnosis and monitoring of complex bipolar disorder, but also provide biological effects of treatment. These assessments devise a new hypotheses about the causes and pathophysiology of bipolar disorder. The peripheral biomarkers, neurotrophins, oxidative stress and neuroinflammation are used to measure the illness activity of bipolar disorder\cite{kapczinski2011peripheral}. The neurotrophins \cite{scola2015role} is a family of proteins that induce the survival, development, and function of neurons. It shows distinct patterns in different stages of bipolar disorder, authors in \cite{fernandes2015peripheral},  use it as brain-derived neurotrophic factor (BDNF) bipolar disorder biomarker. The oxidative stress is an imbalance of antioxidants and free radicals in the human body. Its ratio increases during bipolar disorder episodes therefore in \cite{andreazza2008oxidative} oxidative stress is used to diagnose the bipolar disorder. 
 Neuro inflammation  is a localized physical condition of the body that becomes reddened or swollen during bipolar disorder. The authors of \cite{benedetti2020neuroinflammation} use it for bipolar disorder study. 
The spectral entropy modulation that quantifies the EEG signal degree of uncertainty gets less during  manic episodes of bipolar disorder, so in \cite{vellante2020euthymic} it is used as a bipolar disorder and schizophrenia biomarker for altered function.
The brain oscillations and lithium response variate during bipolar disorder condition and is used as a bipolar disorder biomarker in \cite{atagun2016brain} .

\section{Neural Networks based approaches for Bipolar disorder recognition using EEG signals}
\label{sec:7}
 Bipolar disorder (BD) is a composite disorder that oscillates between two mood states of depression and mania  and is often misdiagnosed by physicians \cite{Ref139}. It is an enduring condition in which BD patients spend most of their life in a misery of symptoms of depression, which complicates accurate identification and analysis of bipolar disorder. The fundamental clinical challenge of this disease is the differentiation between BD patients and the patients showing symptoms of general depression. 
Neural networks (NNs) based approaches offer new ways to predict bipolar disorder recognition and clinical outcomes for individuals,  they reduce the complexity of bipolar disorder identification process.  NNs process the psychiatric data that is relayed on the brain structure and resolve the complex real-world problems which is otherwise difficult for conventional tools and techniques. In contrast to the human data processing ability, NNs provide better precision and time effective solutions for pattern recognition and prediction problems.

Multi-layer perceptron (MLP) is the most popular supervised neural network that is widely used for classification and assessments. It is a class of FFNNs (Feed Forward Neural Networks) that contains at least three layers: an input layer, a hidden layer and an output layer. In \cite{Ref121}, MLP is used as BD (Bipolar Disorder) classifier with five neurons and one hidden layer. To acquire the best conceivable precision, several discriminative features can be extracted from the EEG recordings by using the four diverse feature selection algorithms. These algorithms are CMIM (Conditional Mutual Information Maximization), MIM (Mutual Information Maximization),DISR(Double Input Symmetrical Relevance) and FCBF(Fast Correlation Based Filter) . The extracted features can be fed to the MLP for classification purpose. Results shows that MLP achieve 91.83\% classification accuracy for bipolar disorder sub types and normal subjects \cite{Ref121}. The artificial neural networks and quantitative EEG have been used in \cite{Ref122} for differentiating fronto temporal dementia from late-onset bipolar disorder. All patients are assessed by the clinical MRI scan(Magnetic Resonance Imaging)  and EEG(Electroencephalogram). The results show that a combination of EEG and MRI with ANN classifier gives better classification output as compared to EEG and MRI separately.

As one of the major unbearable neural syndrome, bipolar disorder is commonly misdiagnosed as UD(unipolar disorder) , that further leads to suboptimal cure and poor results. Therefore,the classification of UD and BD at initial phases can therefore support to assist effective and precise treatment. In \cite{Ref40} artificial neural network classifier with quantitative EEG is used as a biomarker for the classification of the unipolar and bipolar disorder and it achieves an accuracy of 89.89\%. A bipolar depression patient experiences both manic and depressive periods. To classify these two depressive periods, the feedforward neural network (FFNN) and probabilistic neural network (PNN) is used in \cite{Ref41}. The FFNN showed 98.75\% classification accuracy while PNN achieved an accuracy of only 46.5\%. The results show that FFNNs give better classification results for bipolar disorder as compared to PNN.

Classification of bipolar EEG signals in normal and depressive condition has been performed in \cite{Ref42} by using relative wavelet energy (RWE) and an artificial feed forward neural network. The performance of the artificial neural network has been assessed by the classification accuracy and its value of 98.11\% shows its unlimited potential for classifying normal and depressive subjects. In \cite{Ref126} convolutional neural network with electroencephalography features is used for the precise diagnosis of the depression sub types. It achieves 99.5\% accuracy in the classification of unipolar vs healthy subjects and 85\% in discrimination of bipolar vs healthy subjects. In \cite{Ref132}, bipolar and schizophrenia disorder diagnosis is achieved by using an artificial neural network. It achieves 90\% classification accuracy among the bipolar, schizophrenia and healthy subjects.

\subsection{EEG Experimental Protocols for Bipolar disorder Recognition}
EEG experimental protocols are a set of rules that are defined before EEG recording. It includes the number of subjects that have participated in the EEG based study, selection criteria of participants, placement standard and types of electrodes that are used for recording the bipolar activity.

\textbf{Participants:}
The ratio of the number of participants for the diagnosis of bipolar disorder varies in different studies according to their resources and requirements. Participants' strength, gender, age group, prior history of medication has a great effect on the diagnosis of bipolar disorder; therefore, most of the studies consider it as their initial protocols requirements as shown in Table \ref{tab:bparticipant}.
As far as participant strength is concerned,
\cite {Ref121,Ref122} include 38 subjects with the age group in the range of 15 to 16 years and 18:20 ratio, which means 18 subjects belongs to the bipolar disorder type I and 20 belongs to the bipolar disorder type II. No distinction about gender is mentioned in this study. To classify the unipolar, bipolar and healthy subjects \cite {Ref40,Ref41,Ref42,Ref125} have performed EEG experiments for 30 ,60, 89 and 134 subjects in which half of the participants are depressed and the others are healthy. Participants age varies from 20 to 50 years with no discrimination of gender. The classification of bipolar, schizophrenia and healthy subjects is performed in \cite{Ref132} in which 35 participants have schizophrenia, 35 are with bipolar illness and the remaining 35 are healthy subjects. From the available literature of bipolar disorder diagnosis, it is observed that there is no standard ratio of number of participants and it varies in different studies according to their resources and requirements of the experiment.
\begin{table*}[]
 \caption{Participants information in EEG experiments for bipolar disorder detection.}
 \begin{center}
 \begin{tabularx}{12cm}{p{1cm}|p{3cm}|p{1cm}|p{2.5cm}|p{4cm}}
    \hline
    Ref & Subjects & Gender & Age Group\newline{(Mean,±SD)}& Bipolarity \\ [0.5ex] 
    \hline
    \cite{Ref121} & 38(18BDI,20BDII) &18F, 20M & 15-16 (15.7,1.5) & Bipolar disorder I, II\\ 
    \hline
    \cite{Ref122} &38(18 bvFTD+20BD) & - & 52-77(64 years) & Bipolar disorder \\
    \hline
    \cite{Ref40} & 89 & - & - & Unipolar, bipolar \\ 
    \hline
    \cite{Ref41} & 60(30D+30H) & 32F, 28M & 20-50& Bipolar depression \\ 
    \hline
    \cite{Ref42} & 30 & 16F, 14M & 20-50& Bipolar depression\\
    \hline
    \cite{Ref125}&134 (75D+59H)&-&18-54(35,4.2) &Bipolar disorder\\
    \hline
    \cite{Ref132}&105 (35SZ+35BD+35H) &-&-& Bipolar disorder\\
    \hline
\end{tabularx}
\end{center}
\label{tab:bparticipant}
\end{table*}
\\

\textbf{Selection Criteria of Bipolar Disorder Participants:}
 The participants are selected based on two major parameters of self-reported psychometric test and inclusion criteria of participants. For all subjects, the inclusion criteria are history of epilepsy, head injury, psychiatric disorders and effect of illegal drugs as shown in Table \ref{tab:bselection}. The Diagnostic and Statistical Manual of Mental Disorders (DSM-IV) and Back Depression Inventory (BDI-II) are the two major self-reported psychometric tests that are used in bipolar studies \cite{Ref41,Ref42,Ref125,Ref132} for primary selection of the participants. Based on inclusion criteria and psychometric test, only those participants are considered for EEG study that have no psychotic disorder and their self-reported psychometric test score is above 14. Interview by a psychiatrist is also conducted in some studies \cite{Ref32} to ensure that the selected participants are taking no medication to treat bipolarity and depression. After the selection of the participants, a ten minutes EEG experiment is performed with open and close eye conditions for each subject.
 \begin{table*}[]
\caption{Participant selection and experimentation tasks in EEG studies for bipolar disorder detection.}
\begin{center}
 \begin{tabularx}{12cm}{p{1cm}|p{3cm}|p{2cm}|p{2cm}|p{2.5cm}}
 \hline
  Ref & EEG Selection Criteria & Subject Inclusion criteria & Tasks& Bipolar Disorder criteria \\ [0.5ex] 
 \hline
  \cite{Ref121} & No psychotic &  DSM-IV,BDI &-& (BDI-II score > 14 \\ 
 \hline
 \cite{Ref122} &  Brain hospital patient & DSM-5 &-& DSM-5 score\\ 
 \hline
 \cite{Ref40} & No mental disorder & & Watch Pics& BDI(II)score 14-28 \\ 
 \hline
 \cite{Ref41} & No mental disorder & BDI-II &-& BDI(II)score 14-28 \\ 
 \hline
 \cite{Ref42} & No neurological history&-& - \\
 \hline
 \cite{Ref125}& Heroin addiction aatients&DSM-IV&Stay relaxed and awake&DSM-IV score\\
 \hline
 \cite{Ref132}& Age no more than 65& DSM-IV&-&-\\
 \hline
 \end{tabularx}
 \end{center}
     \label{tab:bselection}
 \end{table*}

\textbf{Placement and Types of EEG Electrodes:}
The electrode placement and types of electrodes play a major role in EEG based bipolar disorder data acquisition. Minor mistakes in electrode placement pollute the overall EEG results that further affect the classification of the EEG signals. The two major international electrode placement standards at the scalp are 10-10 and 10-20 \cite{seeck2017standardized}. According to the literature presented in Table \ref{tab:bplacement} most of the studies used international placement standards  \cite{Ref121,Ref122,Ref41} but few create their own electrode placement strategy  \cite{Ref40,Ref42}. 
Wet and dry types of electrodes are used in EEG based studies to diagnose bipolar disorder. Wet electrodes are usually made of silver and silver chloride material and applied on a scalp by using the electrolytic gel material that works as a conductor between the skin and the wet electrodes. Dry electrodes consist of a single metal and can be directly placed on the scalp without the need to apply the conductive gel. Dry electrodes are the most efficient and easy to use as compared to wet electrodes. As patients feel comfortable with dry electrodes therefore latest technologies prefer to use them \cite{kam2019systematic}.

\begin{table*}
 \caption{EEG devices,number of channels and their placement used for EEG experiment for bipolar disorder detection.}
 \centering
 \begin{tabularx}{13cm}{p{1cm}|p{2cm}|p{2cm}|p{2cm}|p{2cm}|p{3cm}}
 \hline
  Ref & EEG Device & Electrode & Brain Lobes & Placement Standard & Types of \newline{Electrode} \\
 \hline
  \cite{Ref121} & EEG cap & 19 &-&	10-20 &  Wet  \\ 
 \hline
 \cite{Ref122} &  & 19 & Frontal and temporal  & 10-20 &  Wet  \\
 \hline
 \cite{Ref40} & Procomp & 1(F4) &-& -  &  Dry \\
 \hline
\cite{Ref41} & - & 24 & -&10-20 &  Wet \\
\hline
\cite{Ref42} & - & Two channel pair & Left/right half & - & Wet \\
\hline
\cite{Ref125}& “Neuroscan/ scan LT” neuro-headset&-&-&-&-\\
\hline
\cite{Ref132}&-&-&-&-&-\\
\hline
 \end{tabularx}
     \label{tab:bplacement}
 \end{table*}
 \subsubsection{EEG based public datasets for bipolar disorder.}
 Several public datasets, such as BioGPS \cite{Ref153}, Bipolar Disorder Neuroimaging Database \cite{Ref154}, Bipolar Disorder Phenome Database \cite{Ref155} etc., exist for bipolar disorder recognition. However, after an extensive literature survey, we found that no EEG based public datasets are available for bipolar disorder recognition research.
\subsection{Automatic assessment of Bipolar disorder}
 Automation of bipolar disorder assessment is a major challenge for the research community. The automatic methods make the investigation procedure quick and easy. In these methods, features are automatically extracted from the EEG signals and then based on these features a mental disorder is diagnosed. In this article, artificial neural network and deep learning-based approaches are considered for automated bipolar disorders diagnosis.
\subsubsection{Pre-processing}
The pre-processing or artifact removal is a major part of EEG data acquisition, except this activity no analysis can be performed directly on the EEG data. It is necessary to filter the EEG signals from the different physiological and non-physiological artifacts and interference before using it. The researchers used different techniques and tools to manually or automatically remove the artifacts from the EEG signals as shown in Table \ref{bartifact}.

The band pass, notch and ICA (Independent Component Analysis) are the important filters that are used in most of the bipolar disorder studies \cite{Ref121,Ref122,Ref40,Ref42,Ref125} to remove the noise, interference and physiological and non-physiological artifacts with 40 Hz to 70Hz frequency range. These filters remove the artifacts from EEG signals based on the frequency and amplitude level of the EEG signals (upper and lower level of the signals). The Total Variation Filtering (TVP) and visual inspection are signal filtering techniques that are used in some studies \cite {Ref41} and remove noise manually from the signals.

\begin{table}
\caption{Artifacts and noise filtering approaches in EEG based bipolar disorder detection.}
 \begin{tabularx}{7cm}{p{1cm}|p{5.3cm}}
 \hline
  Ref & Artifact Filtering Technique \\ 
 \hline
  \cite{Ref121} &  Bandpass, notch filter and \newline{visual inspection} \\ 
 \hline
 \cite{Ref122} &  Bandpass filter at 0.15 to 70 Hz \\
 \hline
 \cite{Ref40} & Bandpass filter 40HZ \\ 
 \hline
\cite{Ref41} & Total variation
Filtering (TVP) \\ 
\hline
\cite{Ref42} & Notch filter at 50HZ \\
\hline
\cite{Ref125}& Independent component analysis \\
\hline
\cite{Ref132}&-\\
\hline
 \end{tabularx}
 \label{bartifact}
\end{table}

\subsubsection{Neural network based approaches for Bipolar disorder recognition}
For the past few decades, neural networks are widely used in EEG based studies due to their ability of self-learning and producing the output that is not limited to the input provided to them. Different models of ANN are used in EEG based bipolar disorder diagnosis and classification as shown in Table \ref{tab:bmodel}. The Multi-layer perceptron (MLP), Feed forward neural network (FFNN), Probabilistic neural network (PNN) and Artificial neural network (ANN) are the models that are used in bipolar disorder studies \cite {Ref121,Ref122} with online feature extraction mechanism. They achieve a 91.83\%, 58.75\%, 98.75\% and 90\% classification accuracy respectively. 

\subsubsection{Deep learning based approaches for Bipolar disorder recognition}
Bipolar disorder is often confused with major depressive disorder and other mental disorders. It is characterized by persistent depression and mania. The complications and interruptions in the diagnosis of bipolar disorder delay effective treatment of patients. Bipolar disorder is misdiagnosed as recurrent MDD in 60\% of patients that seek treatment for depression\cite {goodwin2007manic}.To overcome all these limitations and to provide effective diagnosis and treatment to bipolar patients, there is a need to use deep learning based EEG for bipolar disorder diagnosis.

\begin{table}
\caption{Neural Network based approaches for EEG based bipolar disorder detection and their accuracy level.}

 {\begin{tabularx}{7cm}{p{1cm}|p{2.5cm}|p{3cm}}
 \hline
  Ref  & Model used & Results \newline{(Accuracy)} \\ 
 \hline
  \cite{Ref121}&  Multilayer perceptron (MLP) neural network with hidden layer containing 5 neurons &	91.83\% \\ 
 \hline
 \cite{Ref122}& 3-layered ANN with hidden layers containing 20 neurons& 76\% \\
 \hline
 \cite{Ref40}&Back-propagation
learning algorithm containing one hidden layer with 20 neurons & 89.89\%  \\
\hline
\cite{Ref41}&  Two layers Feed forward
neural network (FFNN) and three layers Probabilistic neural network (PNN) &58.75\%, \newline{98.75\%} \\
\hline
\cite{Ref42}& Relative wavelet energy and Feed forward neural network& 98.11\% \\
\hline
\cite{Ref125}& Back-propagation learning algorithm with one hidden layer and 30 neurons &-\\
\hline
\cite{Ref132}& Artificial neural network with six neurons in hidden layers &90\%\\
\hline
\end{tabularx}}
\label{tab:bmodel}
\end{table}

\begin{table}
    \caption{Bipolar disorder type and their criteria defined by DSM-IV.}
    \centering
    \begin{tabularx}{8cm}{p{3cm}|p{4cm}}
    \hline
    Bipolar Disorder Type& (DSM-IV) criteria  \\
    \hline
    Bipolar disorder type I& In bipolar disorder type I one full episode of mania or diverse episodes of mania and depression can occurs. \\
    \hline
    Bipolar disorder type II & In bipolar disorder type II ,no manic episode, minimum one hypomanic episode and many depressive episodes occurs.\\
    \hline
    Cyclothymic disorder& Many hypomanic and depressive episodes occurs. \\
    \hline
    Bipolar disorder not otherwise specified &Depressive and hypomanic episodes  changes rapidly.\\
    \hline
    \end{tabularx}
    \label{tab:my_label}
\end{table}

\section{Discussion}
\label{sec:8}
Our umbrella review provided an up-to-date overview and synthesis of the literature of neural networks-based approaches using EEG signals for the diagnosis of Major Depressive Disorder and Bipolar Disorder. We included data from different sources of Healthy Brain Network (HBN),  PREDICT, pubMed, IEEE explore, embase, google scholar, research gate and web of science etc.  We have noticed  several limitations, which have been discussed with improvement suggestions as follows:\\

\textbf{Few Existing Works:} Deep learning approaches for EEG based depression diagnosis have been found in only 4 to 5 recent articles from 2018-2020 \cite{Ref31,Ref34,Ref32}.
In case of EEG based bipolar disorder no deep learning based article exists for diagnosis. 
Despite the few existing works in deep learning-based approaches for EEG depression and bipolar disorders analysis, deep learning seems to be promising in many research areas. It can play a key role in developing a more accurate biomarkers. 

\textbf{Data Availability:} The significant issue regarding  the data is its availability. Most of the techniques reviewed here are on private datasets, very few public datasets exist for EEG based depression diagnosis due to the sensitive nature of depression data, and for privacy and confidentiality reasons as well. In bipolar case no public dataset is available for EEG based bipolar disorder diagnosis.
However, for collective innovation, there should be some standard public datasets. Open datasets provide new opportunities for collaborations. Further, these datasets can be useful to evaluate and validate methodologies/approaches presented by different researchers with standardized protocols and settings.

\textbf{AI for a psychiatry revolution:} EEG is a non-invasive strong biomarker, mental health and psychiatry do not benefit as yet.
Suggestion:  This domain needs to attract more attention from the scientific community especially the computer vision researchers with more innovative methods and applications for better diagnosis of mental disorders. 

\textbf{ Deep learning for more complex patterns:} The feature extraction phase is very important in any computer vision pipeline and participate strongly in the precision and accuracy level of the methodology. The classification is applied only on extracted features. In fact, the majority of the approaches reviewed here used the conventional machine learning approaches for feature extraction that does not predict how many  features should be extracted for a high classification accuracy.
Suggestion: Feature extraction phase should be automatic and predictable for higher classification ratio as in deep learning-based approaches. 

\textbf{Need for Multimodal approaches:} The multimodal approaches of EEG and its fusion with other modalities are less explored in literature, despite the fact that multimodal fusion performs better than unimodal approaches in many other applications \cite{Ref157}.
Suggestion: EEG based multimodel deep learning approaches should be introduced in depression and bipolar disorder detection to enhance the classification accuracy.

\textbf{ EEG signals and noise:} Most of the studies reviewed here use conventional artifacts removal techniques likewise: visual inspection, notch filter, low band and high pass filter to sanitize the polluted EEG signals. These conventional approaches cannot filter the signals accurately and signals still carry a lot of noise, which yield a negative impact on depression detection/ classification accuracy \cite{Ref158}. 
Suggestion: The signal pre-processing or filtering can be automated by using more advanced techniques like deep learning approaches. 

\textbf{ Deep learning and interpretability:} The problem of interpretability of the results of neural networks based approaches still exist. These neural networks-based approaches can be regarded as black boxes. Without the interpretability of these algorithm results, it is difficult for a physician to tell patients about their predicted diagnosis with any genuine reason. Decision-making in such uncertain situations is a problem of liability, ethics and beyond pure performance. Therefore, the final decision of algorithm should be interpretable \cite{Ref163}. 
Suggestion: The neural networks-based results should be interpretable to enhance the understanding level and able to work in more certain scenarios.

\textbf{Brain Lobes:} More than 90\% of reviewed studies use the frontal lobe and left/right hemisphere for depression and bipolar disorder detection as given in Table \ref{tab:electrode}, and achieve a remarkable classification accuracy. It can be a strong motivation for the new researchers for future analysis on frontal lobe.

\section{Conclusion}
\label{sec:9}
Mental disorders are highly prevalent and disabling health condition. Numerous studies explored the use of EEG signals to diagnose the functioning of brain activity. In this survey paper, the focus is giving to EEG signals as a strong biomarker for Major Depressive Disorder and Bipolar Disorder. And thus, an extensive study of the state-of-the-art shallow and deep neural networks based methods is giving for MDD and BD diagnosis and assessment. While EEG based methods for MDD diagnosis attracts the attention of the computer vision community, the EEG based methods for BD diagnosis is less explored in the literature and need more consideration. The EEG based experimental protocols and methods could help the scientific community to better understand mental disorders and to design strong biomarkers for their diagnosis and assessment.\\
Deep neural networks offer high classification accuracy among the depressed and healthy control subjects in comparison to shallow neural networks based methods. Several clinical research issues remain to be addressed scientifically in this field. Thus, a set of recommendations are addressed in the discussion section to offer ways to guide the emerging collaborations and interactions towards the most fruitful outcomes.\\
In conclusion, the current exhaustive review highlights the considerable potential of the EEG-based methods for the assessment and monitoring of MDD and BD. However, despite the good prediction performance of neural networks based methods, they lack sufficient model explainability needed in quantitative research and which prevent the community to further develop reproducible and deterministic protocols and to achieve clinically useful results. Thus, explainable neural networks based methods are needed for mental health diagnosis and assessment. Finally, we hope this paper is another step towards harnessing the full potential of AI for mental health diagnosis. 

\section*{Declaration of Competing Interest}
The author(s) declare(s) that there is no conflict of interest.

\bibliographystyle{elsarticle-num}
\bibliography{Ref}




\end{document}